\documentclass[twoside,11pt]{article}

\usepackage{obs_study_style}
\usepackage{amsmath}    
\usepackage{bm}          
\usepackage{listings}    
\usepackage{booktabs}    
\usepackage{enumitem}    
\usepackage{algorithm}
\usepackage{algorithmic}
\usepackage{setspace}

\usepackage[colorlinks=true,citecolor=blue,urlcolor=blue,linkcolor=blue]{hyperref}

\newcommand{\iid}{\stackrel{iid}{\sim}}
\newcommand{\indep}{\stackrel{indep}{\sim}}
\newcommand{\true}{{\mbox{\scriptsize tr}}}
\newcommand{\Nor}{\mbox{N}}

\newcommand{\Ber}{\mbox{Ber}}
\newcommand{\Unif}{\mbox{Unif}}
\newcommand{\Dat}{\mbox{${\cal D}$}}

\newtheorem{result}[theorem]{Result}

\heading{}{}{}{}{}{Yongseok Hur, Joonhyuk Jung and Juhee Lee}

\ShortHeadings{Causal Logistic Stick-Breaking Process}{Hur, Jung and Lee}
\firstpageno{1}

\begin{document}

\title{A General Bayesian Nonparametric Approach for Estimating Population-Level and Conditional Causal Effects}

\author{\name Yongseok Hur \email 628hur@snu.ac.kr \\
       \addr Department of Statistics\\
       Seoul National University\\
       Gwanak-ro 1, Gwanak-gu, Seoul 08826, Republic of Korea
       \AND
       \name Joonhyuk Jung \email joonhyukjung@uchicago.edu \\
       \addr Department of Statistics\\
       University of Chicago \\
       5747 S. Ellis Avenue, Chicago, IL 60637, USA
       \AND
       \name Juhee Lee \email jle297@ucsc.edu \\
       \addr Department of Statistics\\
       University of California, Santa Cruz \\
       1156 High Street, Santa Cruz, CA 95064, USA}

\maketitle

\begin{abstract}
We propose a Bayesian nonparametric (BNP) approach to causal inference using observational data consisting of outcome, treatment, and a set of confounders. The conditional distribution of the outcome given treatment and confounders is modeled flexibly using a dependent nonparametric mixture model, in which both the atoms and the weights vary with the confounders. The proposed BNP model is well suited for causal inference problems, as it does not rely on parametric assumptions about how the conditional distribution depends on the confounders. In particular, the model effectively adjusts for confounding and improves the modeling of treatment effect heterogeneity, leading to more accurate estimation of both the average treatment effect (ATE) and heterogeneous treatment effects (HTE). Posterior inference under the proposed model is computationally efficient due to the use of data augmentation. Extensive evaluations demonstrate that the proposed model offers competitive or superior performance compared to a wide range of recent methods spanning various statistical approaches, including Bayesian additive regression tree (BART) models, which are well known for their strong empirical performance. More importantly, the model provides fully probabilistic inference on quantities of interest that other methods cannot easily provide, using their posterior distributions.
\end{abstract}


\section{Introduction}

Understanding causal relationships from observational data is a cornerstone of informed decision-making in various fields including policy design and scientific discovery \citep{imbens2015causal, lunceford2004stratification}. Unlike data from randomized controlled trials, which can be costly, ethically challenging, or impractical, observational data are often readily available at large scale \citep{yao2021survey}. While the accessibility of observational data is appealing, it necessitates careful analysis due to the presence of missing counterfactuals and selection bias.

Among many causal effects that may be of interest to researchers, the average treatment effect (ATE), which quantifies the overall impact of an intervention on a population, is often the primary target of inference. Established methodologies for ATE estimation include regression adjustment, which models the outcome as a function of treatment and covariates \citep{hill2011bayesian}; matching techniques that aim to create comparable treated and control groups based on observed characteristics \citep{stuart2010matching, linden2016using}; and weighting methods that adjust for imbalances in covariate distributions between treatment groups \citep{imai2014covariate, hainmueller2012entropy}; and doubly robust methods that combine outcome regression and propensity score modeling to achieve consistency if either model is correctly specified \citep{robins1994estimation}. Recently, for applications such as personalized medicine, where treatments are tailored to individual patient characteristics, and targeted marketing, where interventions are optimized for specific customer profiles \citep{nie2021quasi, kennedy2022towards, lei2021conformal}, the estimation of 
heterogeneous treatment effects (HTE) has received increasing attention. Estimating HTE means assessing how the causal effect varies across different segments of of the population, or across individuals; in practice this is often formalized by estimating conditional average treatment effects (CATE) as functions of covariates to capture effect heterogeneity. 
%
However, the estimation of HTE presents substantial challenges due to the presence of confounding variables that can obscure the true causal effect, the potential for complex relationships between covariates and outcomes, and the complexity arising from interactions among multiple covariates. Different estimation methods can produce markedly different HTE estimates, underscoring the need for great care in both the choice of methodology and the interpretation of the results \citep{kurth2006results, lunt2009different}.  

Recent advances in Bayesian causal inference, particularly those using Bayesian nonparametric approaches, have provided promising avenues to address the complexities involved in estimating both ATE and HTE \citep{vegetabile2020optimally, liu2020estimation, xu2018bayesian, hahn2020bayesian}. Bayesian methods offer a principled and straightforward means of uncertainty quantification and incorporating prior knowledge. In particular, Bayesian additive regression tree (BART) models have gained significant popularity due to their strong empirical performance \citep{hill2020bayesian, hahn2020bayesian}. Its success has spurred a number of follow-up studies and extensions, e.g.,  \cite{krantsevich2023stochastic, kim2023bayesian} among many others. In addition, Bayesian nonparametric approaches based on Dirichlet process mixtures and Gaussian processes have been successfully applied to causal effect estimation problems (see, for example, \citet{roy2018bayesian, ray2019debiased, zorzetto2024confounder}). The effectiveness of these Bayesian approaches can still be influenced by the specific structure of the data and the degree of confounding present \citep{oganisian2021practical, ding2023posterior, linero2023prior}. Comprehensive reviews and classifications of these Bayesian methodologies can be found in \citet{stephens2022causal, li2023bayesian, linero2023and}. 

In this paper, we adopt a flexible Bayesian approach using a dependent nonparametric mixture model to build an outcome regression framework for general causal inference. Dependent Dirichlet processes (DDPs) \citep{maceachern2000dependent, quintana2022dependent}, which define a flexible class of predictor-dependent random probability distributions, along with many of their extensions, have been successfully applied in various domains. Specifically, we construct independent weight and atom processes indexed by covariates, similar to the DDP framework, allowing the distribution of the outcome to vary flexibly with the confounders. In particular, the weights of the mixture components are modeled using the logistic stick-breaking process (LSBP), originally introduced by \citet{ren2011logistic} for nonparametric clustering. Estimated propensity scores are incorporated as an additional covariate, thereby enhancing the model’s ability in estimating treatment effects \citep{hahn2018regularization, oganisian2025priors}. Unlike tree-based models, the proposed method adaptively models covariate-dependent heterogeneity through continuous mixing distributions and can capture smooth, continuous variations in treatment effects as well as abrupt distributional changes \citep{wade2023bayesian}, which enables efficient inference of complex heterogeneity in treatment effects. Furthermore, we leverage tractable posterior simulation techniques using the augmentation of P\'olya-Gamma variables \citep{polson2013bayesian, rigon2020tractable} to achieve efficient posterior computation.  
While the proposed model has been studied in various contexts, to our knowledge, it has not been thoroughly examined in causal inference applications. To benchmark its effectiveness, we evaluate the model’s performance against a broad range of recent methods, with particular emphasis on comparisons to BART-based models, using extensive simulation studies, a reanalysis of the 2022 ACIC Data Challenge data, and a real data application. The proposed model demonstrates overall competitive performance, showing particularly strong results in estimating HTEs, especially in challenging scenarios with small subgroup sizes. In the real data analysis, we illustrate how the model can be used to infer various causal quantities, including quantile treatment effects, which are not straightforward to estimate using tree-based methods. We also show that the model provides fully probabilistic inference on quantities of interest that other methods cannot easily provide through their posterior distributions.

The remainder of this paper is structured as follows. Section 2 provides a detailed overview of related work and presents the development of the proposed methodology. Sections 3 and 4 report on simulation studies, reanalyses of the 2022 ACIC Data Challenge data, and an analysis of a 401(k) pension dataset. Section 5 concludes with a discussion and directions for future work.

\section{Statistical Model for Causal Inference}
\subsection{Assumptions and Estimands}

Following the potential outcome framework \citep{rubin2005causal}, we let $Z \in \{0, 1\}$ denote a binary treatment variable, and $Y(Z)$ the potential outcomes that would have been observed under treatment $Z$. Also, we let $X$ represent a collection of potential confounders. We then observe the potential outcome for the realized treatment;
\begin{align*}
  Y^{\mathrm{obs}} = Y(Z) = ZY(1) + (1-Z)Y(0). 
\end{align*}
Only one of $Y(0)$ and $Y(1)$ is observed for each unit. For brevity, $Y$ will be used to represent $Y^{\mathrm{obs}}$.
Throughout the paper, we assume the stable unit treatment value assumption (SUTVA) \citep{rubin1980randomization} that only two potential outcomes exist and one of them is observed for each subject and that the outcome for any given unit (e.g., a person or subject) should not be affected by the treatment assigned to other units. In addition, we assume ignorability and positivity;
\begin{align*}
  (Y(1), Y(0)) \perp\!\!\!\!\perp Z \mid X \tag{ignorability} \\
  \pi(X) := \mathbb{P}(Z = 1 \mid X) \in (0, 1) \tag{positivity}
\end{align*}
Here, $\pi(X)$ denotes the propensity score, the conditional probability of treatment assignment given $X$ \citep{rosenbaum1983central, rosenbaum2023propensity}. The ignorability assumption states that the potential outcomes are conditionally independent of the treatment assignment given control variables.
Positivity, also called overlap, ensures that a subject has a non-zero probability of receiving either treatment, and there are no groups of subjects for whom the treatment assignment is deterministic.The ignorability assumption together with overlap, is called strong ignorability. The assumptions, while untestable, are required for identifying the causal estimands.

Under the aforementioned assumptions, we consider the problems of estimating population average treatment effect(PATE, often simply called ATE), $\tau$, and conditional average treatment effect (CATE, $\tau(X)$);
\begin{align}
  \tau &= \mathbb{E}\left(Y(1) - Y(0)\right) = \mathbb{E}(\tau(X)), \label{eq:PATE} \\  
  \tau(X) &= \mathbb{E}\left(Y(1) - Y(0) \mid X\right).  \label{eq:CATE}
\end{align}
The CATE $\tau(X)$ is the average effect of a treatment for a specific subgroup of the population defined by certain covariates. The PATE $\tau$ averages these conditional average treatment effect over the covariate distribution. Estimating the population average treatment effect in \eqref{eq:PATE} requires marginalizing $\tau(X)$ over the distribution of $X$ in target population.
In practice, one often approximates this population-level estimand by averaging $\tau(X)$ over the empirical distribution of $X$, yielding the mixed average treatment effect (MATE) \citep{li2023bayesian}, 
\[
\tau^M = \frac{1}{n}\sum_{i=1}^n \tau(X_i),
\] 
where $\{X_i, i=1, \ldots, n\}$ are the observed values of the confounders for $n$ subjects drawn from the target population.
The CATE characterizes heterogeneous treatment effects across covariate profiles, while the MATE serves as the empirical analogue of the PATE obtained by averaging the conditional effects over the observed covariate distribution. For clarity, throughout the rest of the paper, the estimand reported as ‘ATE’ refers to this empirical analogue unless otherwise noted.

\subsection{Model Specification}

To estimate a causal effect from the observed data $(Y, Z, X)$, we build a conditional model for $Y$ given $Z$ and $X$;
\begin{align*}
    Y &\mid Z, X \sim f(Y \mid Z, X). 
\end{align*}
Assuming a predetermined form for $f(Y \mid Z, X)$ can potentially limit the ways $X$ and $Z$ affect the distribution of $Y$. By utilizing a Bayesian nonparametric approach that offers flexible models for conditional distributions, we use a dependent nonparametric mixture model and let 
\begin{align}
    \begin{split}
    f(\cdot \mid Z, X) &= \int \mathcal K(\cdot \mid \phi) \mathrm dG_{Z, X}(\phi),     \label{eq:DDP} \\
    G_{Z, X} &= \sum_{h=1}^\infty w_h(X)\delta_{\theta_h(Z, X)}.  
\end{split}
\end{align} 
Here, $\mathcal K(\cdot \mid \phi)$ is a given parametric kernel indexed by $\phi$. $G_{Z, X}$ is the mixing distribution that is a discrete distribution with atoms $\theta_h(Z, X)$ and weights $w_h(X)$, where $w_h(X)$ are covariate-dependent weights constructed via a stick-breaking representation, and $\theta_h(Z, X)$ are independent stochastic processes with $Z$ and $X$.
The model in \eqref{eq:DDP} allows the entire shape of $f(\cdot \mid Z, X)$ to change with $Z$ and $X$  and builds richer support than models with constant weights \citep{rodriguez2011nonparametric, wade2023bayesian}.
We use a normal distribution $\Nor(\mu_h(X) + \tau_h(X)Z, \sigma^2_h)$ for the kernel in \eqref{eq:DDP} and incorporate $X$ into $\mu_h$ and $\tau_h$ using regression by letting $\mu_h(X) = \widetilde \beta^\top_h \phi_\beta(X)$ and $\tau_h(X) = \widetilde \gamma^\top_h \phi_\gamma(X)$; 
\begin{eqnarray}
Y \mid G_{Z, X} \sim \sum_{h=1}^\infty w_h(X) \Nor\left(\widetilde \beta^\top_h \phi_\beta(X) + \widetilde \gamma^\top_h \phi_\gamma(X) Z, \sigma^2_h\right), \label{eq:DDP-mixture}    
\end{eqnarray}
where $\phi_\beta(X)$ and $\phi_\gamma(X)$ denote collections of selected functions of the covariates.  Nonlinear functions of $X$ can be incorporated in $\phi_\beta(X)$ and $\phi_\gamma(X)$ for additional flexibility (e.g., \cite{rodriguez2025density}). 
We let $\phi_\beta(X)$ and $\phi_\gamma(X)$ include propensity score estimate $\hat{\pi}(X_i)$.
Conditioning on the propensity scores enables an effective estimation of the treatment effect, $\tau(X)$, even when there is confounding and the model for predicting $Y$ based on $Z$ and $X$ is misspecified \citep{rosenbaum1983central}. Furthermore, including $\hat{\pi}(X_i)$ as a covariate in the model reflects prior beliefs about the degree of selection bias and can reduce bias in treatment effect estimates in finite samples \citep{hahn2020bayesian, linero2023prior}.
Under \eqref{eq:DDP-mixture}, CATE and MATE can be estimated by $\tau(X) = \sum_{h=1}^\infty w_h(X)\tau_h(X)$ and $\tau^M = 1/n\sum_{i=1}^n \sum_{h=1}^\infty w_h(X_i)\tau_h(X_i)$, respectively.  In addition, the model can be used to infer various other causal effect quantities, as illustrated in Section~\ref{sec:data-analyses}.

Next, we build prior models for $(\widetilde \beta_h, \widetilde \gamma_h, \sigma^2_h)$ and $w_h(X)$. We let $\beta_h = (\widetilde \beta_h, \widetilde \gamma_h)$ a $p$-dim vector.
We consider an inverse-gamma prior for $\sigma^2_h$, and assuming conditional independence among the elements in $\beta_h$, we consider a horseshoe prior; let $\mathbf \Lambda^{-1}_\beta = \xi^2\mathrm{diag}(\lambda^2_1, \ldots, \lambda^2_p)$.  We then assume the following prior distribution for $(\beta_h, \sigma^2_h)$,
\begin{eqnarray}
\begin{aligned}
&   p(\beta_h, \sigma^2_h \mid \nu_0, s^2_0, \mathbf \Lambda_{\beta}) \propto
        \frac{1}{\sigma_h^{\nu_0 + p + 2}}\exp\left(-\frac{\nu_0 s_0^2 + \beta_h^\top \mathbf \Lambda_\beta \beta_h
        }
    {2\sigma^2_h}\right),  \label{eq:HS} \\
&   \lambda_j \iid C^+(0, 1), j=1, \ldots, p, \mbox{  and  } \xi \sim C^+(0, 1). 
\end{aligned}
\end{eqnarray}
The hyperparameters $\nu_0$ and $s^2_0$ respectively represent the degree of freedom and scale parameter of the inverse gamma prior of $\sigma^2_h$. Here, $\lambda_j$ controls local shrinkage for individual components of $\beta_h$, while $\xi$ governs global shrinkage. By integrating out $\lambda_1^2, \ldots, \lambda_p^2$, the marginal prior of $\beta_h$ corresponds to the horseshoe prior distribution \citep{carvalho2010horseshoe}. 
The horseshoe prior, a regularizing prior, can robustly address the challenge of unknown sparsity in high-dimensional settings.  Furthermore, the inclusion of $\hat{\pi}(X_i)$ in $\phi_\beta(X)$ and $\phi_\gamma(X)$ helps mitigate regularization-induced confounding (\cite{hahn2018regularization}) that may arise from the imposed regularization.

For $w_h(X)$, we consider a logit stick-breaking process (LSBP) \citep{ren2011logistic} and construct the weights via sequential logistic regressions; 
\begin{align}
    w_1(X) &= \frac{\exp(\frac{1}{2} \psi(X)^\top b_1)}{2 \cosh (\frac{1}{2} \psi(X)^\top b_1)}, \nonumber \\
    ~~\mbox{ and }~~
    w_h(X) &= \frac{\exp(\frac{1}{2} \psi(X)^\top b_h)}{2 \cosh (\frac{1}{2} \psi(X)^\top b_h)} \prod_{k=1}^{h-1}\frac{\exp(-\frac{1}{2} \psi(X)^\top b_k)}{2 \cosh (\frac{1}{2} \psi(X)^\top b_k)}, ~ h \geq 2,
    \label{eq:LSBP-w}
\end{align}
where $\cosh(a) = \frac{1}{2}(e^a+e^{-a})$. $\psi(X)$ denotes a collection of $q$ selected functions of $X$. It may be different from $\phi_\beta(X)$ and $\phi_\gamma(X)$, and may include $\hat{\pi}(X)$. We assume $b_h \mid \mathbf \Lambda_b \iid \Nor(0, \mathbf \Lambda_b^{-1})$, where we let $\mathbf \Lambda_b^{-1} = \zeta^2 \mathrm{diag}(\rho^2_1, \ldots, \rho^2_q)$ and assume inverse-gamma priors, independently for $\zeta^2, \rho^2_1, \ldots, \rho^2_q$. 
To prevent the stick-breaking prior probabilities from becoming overly skewed toward 0 or 1, we used an inverse-gamma prior as the hyperprior instead of a half-Cauchy prior in this case.
Covariate-dependent weights can be constructed in various ways (e.g., \cite{griffin2006order, dunson2008kernel, rodriguez2011nonparametric}). Compared to other methods, the LSBP provides a significant advantage in posterior computation by leveraging the Pólya-Gamma data augmentation \citep{polson2013bayesian}. We will discuss this further in the following subsection.

\subsection{Posterior Computation}
\label{sectionposterior}
We implement posterior inference via Markov chain Monte Carlo (MCMC) simulation. In particular, we use a truncated version, $G^H_{Z, X}$, of LSBP in \eqref{eq:LSBP-w} with $(H + 1)$ sticks by letting $w_{H+1} = 1 - \sum_{h=1}^H w_h$ as an approximation of the infinite process. With a sufficiently large value of $H$, the truncated LSBP can provide an accurate approximation of the infinite representation. See Theorem 1 of \citet{rigon2020tractable} for detailed discussion on the convergence of $G^H_{Z, X}$ to $G_{Z, X}$ as $H \rightarrow \infty$. 

Suppose we have data $\Dat=\{\Dat_i, i=1, \ldots, n\}$ with $\Dat_i=(Y_i, Z_i, X_i)$ from $n$ independent subjects. We introduce latent configuration variables $u_i \in \{1, \ldots, H + 1\}$ such that $p(u_i = h \mid X_i, \mathbf b) = w_h(X_i)$, where $\mathbf b = (b_h)_{h=1}^H$. Given $u_i$, we assume $y_i \mid u_i=h, \beta_h, \sigma^2_h \sim \Nor\left(\beta^\top_h \phi(Z_i, X_i), \sigma^2_h\right)$ with $\phi(Z_i, X_i) = (\phi_\beta(X_i)^\top, Z_i \phi_\gamma(X_i)^\top)^\top$. We obtain the joint posterior distribution; 
{\small 
\begin{eqnarray}
\begin{aligned}
    p(\bm u, \bm \beta, \bm \sigma^2, \bm b, \xi^2, \bm{\lambda}^2, \zeta^2,   \bm{\rho}^2 \mid \Dat) 
    &\propto
    \prod_{i=1}^n p(Y_i \mid u_i, X_i, Z_i) 
    \prod_{i=1}^n p(u_i \mid X_i, \mathbf b) 
    \prod_{h=1}^H p(b_h \mid \mathbf \Lambda_b) 
    \label{eq:joint-post}\\ 
    & \hspace{0.2in} \times
    \prod_{h=1}^{H+1} p(\beta_h, \sigma^2_h \mid \nu_0, s^2_0, \mathbf \Lambda_\beta)
    p(\xi^2, \bm{\lambda}^2, \zeta^2,   \bm{\rho}^2).     
\end{aligned}
\end{eqnarray}
}
We exploit  a P\'olya-gamma data augmentation  and facilitate the steps of updating $u_i$ and $b_h$ \citep{rigon2020tractable}. Specifically, we introduce binary random variables, $\eta_{i,h} \in \{-1/2, 1/2\}$, and let 
\begin{align*}
    p\left(\eta_{i,h} = \frac{1}{2} \mid X_i, b_h \right) = \frac{\exp(\eta_{i,h} \psi(X_i)^\top b_h)}{2\cosh(\frac{1}{2}\psi(X_i)^\top b_h)}.
\end{align*}
Observe that we have 
$
    u_i = \min \left(\left\{1 \leq h \leq H: \eta_{i,h} = 1/2 \right\} \cup \{H+1\}\right).
$
Also, we introduce another set of latent variables $\omega_{i,h}$ from a P\'olya-Gamma distribution with parameters 1 and  $\psi(X_i)^\top b_h$, independently of $\eta_{i,h}$'s.

\begin{result}[Full Conditional Distributions]
\label{propposterior}
    Let $\Dat^\star_h = \{(Y_i, \phi(Z_i, X_i))\mid u_i = h\}$ and $\overline{\Dat}^\star_h = \{(\eta_{i,h}, \omega_{i,h}, \psi(X_i)) \mid u_i \geq h\}$, $h=1, \ldots, H+1$. Then, the full conditionals of $(\beta_h, \sigma^2_h)$ and $b_h$ are given as follows;
    \begin{align}
        \label{eqthetaposterior}
        p(\beta_h, \sigma^2_h \mid \Dat^\star_h, \nu_0, s^2_0, \mathbf \Lambda_{\beta}) 
        &\propto  
        \frac{1}{\sigma_h^{\nu_h + p + 2}}\exp\left(-\frac{\nu_h s_h^2 + (\beta_h - \mathbf m_{\beta_h})^\top \mathbf \Lambda_{\beta_h} (\beta_h - \mathbf m_{\beta_h})
        }
        {2\sigma^2_h}\right), \\
        \label{eqbposterior}
        b_h \mid \overline{\Dat}^\star_h, \mathbf \Lambda_b &\sim \mbox{N}\left(\mathbf m_{b_h}, \mathbf \Lambda^{-1}_{b_h}\right),
    \end{align}
    where
    \begin{align*}
        \mathbf \Lambda_{\beta_h} &= \mathbf \Lambda_\beta + \sum_{i=1 \mid u_i = h}^n \phi(Z_i, X_i)\phi(Z_i, X_i)^\top,  
        \hspace{0.2in}
        \mathbf m_{\beta_h} = \mathbf \Lambda_{\beta_h}^{-1}\sum_{i=1 \mid u_i = h}^n Y_i \phi(Z_i, X_i), \\
        \mathbf \Lambda_{b_h} &= \mathbf \Lambda_b + \sum_{i=1 \mid u_i \geq h}^n \omega_{i,h} \psi(X_i) \psi(X_i)^\top, 
        \hspace{0.2in}
        \mathbf m_{b_h} = \mathbf \Lambda_{b_h}^{-1}\sum_{i=1 \mid u_i \geq h}^n \eta_{i,h} \psi(X_i)^\top, \\
        \nu_h &= \nu_0 + \mid\Dat^\star_h\mid,
        ~~~\mbox{ and  }~~~
        \nu_h s^2_h = \nu_0 s^2_0 + \sum_{i=1 \mid u_i = h}^n Y_i^2 - \mathbf m_{\beta_h}^\top \mathbf \Lambda_{\beta_h} \mathbf m_{\beta_h}.
    \end{align*}
\end{result}
\noindent
Due to the conditional conjugacy, $\beta_h$ and $\sigma^2_h$ can be easily updated via the Gibbs sampler as in \eqref{eqthetaposterior}. Furthermore, the full conditional of $b_h$ is a normal distribution conditional on $\eta_{i, h}$ and $\omega_{i, h}$. Rue's algorithm \citep{rue2001fast} is further employed to efficiently sample $\beta_h$ and $b_h$ from their full conditional distributions. Details of the posterior computation are provided in \ref{appendixa} and \ref{appendixb}.

\section{Simulation Studies}
We conducted two simulation studies to evaluate the performance of the proposed causal LSBP model (referred to as cLSBP) and to compare it with that of the Bayesian causal forest (BCF), a Bayesian nonparametric regression model based on BART \citep{hahn2020bayesian}. For Simulation 1 in Section~\ref{simulation1}, we adapted the simulation scenarios from \citet{kang2007demystifying} with minor modifications and set up five scenarios with varying strengths of targeted selection. For Simulation 2, we used the simulation scenarios in Section 6.1 of \cite{hahn2020bayesian} and reported the results in Section~\ref{sec:simulation2}. 

To fit cLSBP, we specified $\nu_0 = 10$ and $s_0^2 = 0.2$ for the prior of $\sigma_h^2$ in all examples for Section~\ref{simulation1}, ~\ref{sec:simulation2}, and ~\ref{sec:ACIC}. For Section~\ref{sec:pension}, we used $\nu_0 = 20$ and $s_0^2 = 0.025$ due to distinctive distributional features of the outcome. We note the prior sensitivity for $\sigma_{h}^{2}$ with details discussed in Section~\ref{sec:pension}.

We specified as $\nu_0 = 10$ and $s_0^2 = 0.2$ for the Section~\ref{sec:pension}, to  While $\phi_\beta(X)$, $\phi_\gamma(X)$, and $\psi_\beta(X)$ could be specified as flexible functions, we set them to the linear identity in all examples:
$
\phi_\beta(X) = \phi_\gamma(X) = \psi_\beta(X) = X.
$
We discarded the first 4000 iterates for burn-in and kept the next 4000 iterates for posterior inference. We examined the convergence and mixing of the MCMC algorithms using standard diagnostic techniques and found no evidence of practical convergence problems. We also performed prior sensitivity analysis by varying the fixed hyperparameter values in a reasonable range. We observed only minimal changes in the resulting inferences.

We employed four comparators: cLSBP with and without propensity score estimates $\hat{\pi}_i = \hat{\pi}(X_i)$, and BCF with and without $\hat{\pi}_i$. Specifically, the propensity function was estimated using BART \citep{chipman2010bart,BART2024package}, and the resulting estimates $\hat{\pi}_i$ were included a covariate in addition to $X$. 
Incorporating $\hat{\pi}_i$ may affect the estimation of treatment effects, especially when strong confounding is present. For BCF, we used R package \textit{bcf} with their default setting. For each method, we estimated the (sample) ATE and CATE and used them to evaluate the root mean square error (RMSE), coverage (COV), and average interval length (LEN).

\subsection{Simulation 1}\label{simulation1}

\begin{figure}[!t]
    \centering
    \includegraphics[width=0.95\linewidth]{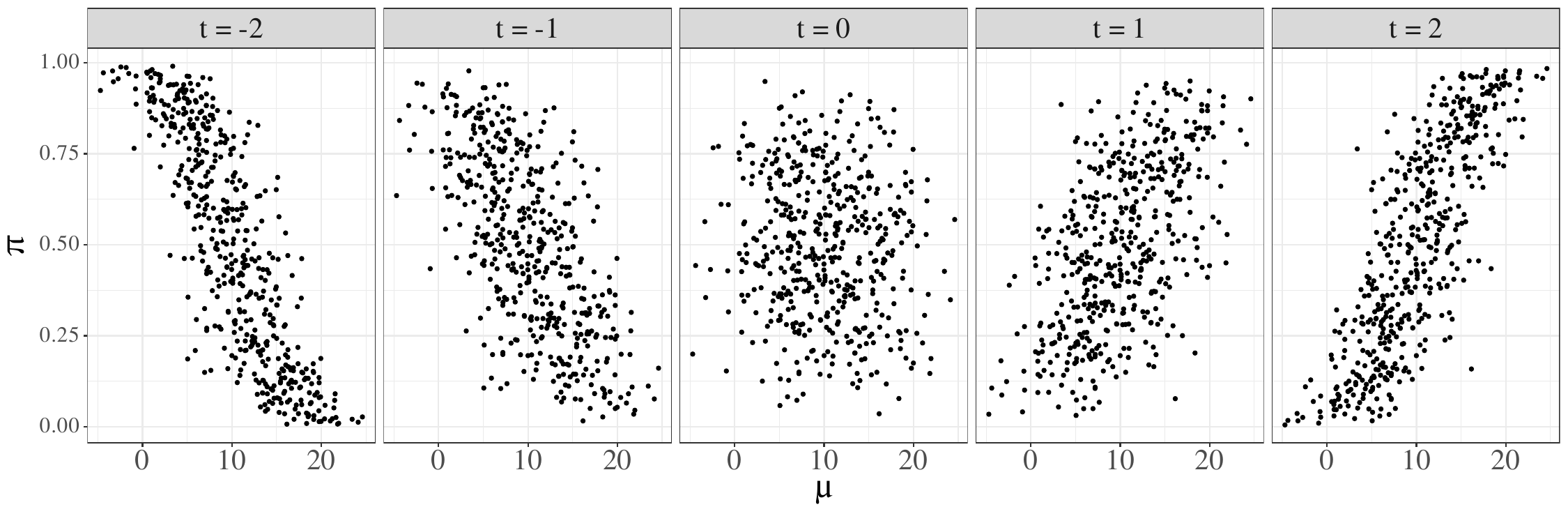}
    \caption{[Simulation 1] Scatterplots of $\pi^\true_i$ versus $\mu_i^\true$ with different values of $t=-2, -1, 0, 1$ and $2$. Here, $t$ controls the strength of targeted selection.}
    \label{fig:Kang_mu_pi}
\end{figure}

Similar to the simulation setup in \cite{kang2007demystifying}, we first simulated latent variables, $W_i = (W_{i1}, W_{i2}, W_{i3}, W_{i4})^\top \overset{iid}{\sim} \mathrm{N}(0, \mathbf I_4)$, $i=1, \ldots, n$ with $n=500$ and specified the true propensity scores ($\pi^\true_i)$, mean without treatment ($\mu^\true_i$), and treatment effect ($\tau^\true_i$) as follows;
\begin{align}
    \pi^\true_i &= \Phi\left(- W_{i1} + 0.5W_{i2} -0.25W_{i3}-0.1W_{i4} + \frac{37}{280}\left(t+1\right)(\mu^\true_i - 10) \right), \label{eq:sim1-pi}\\
    \mu^\true_i &= 10 + 4W_{i1} + 2W_{i2} + 2W_{i3} + 2W_{i4}, \nonumber \\
    \tau^\true_i &= 1 - W_{i2} + 2W_{i4}. \nonumber
\end{align}
We then generated $Z_i \mid \pi^\true_i \indep \Ber(\pi^\true_i)$ and  $Y_i=\mu^\true_i + \tau^\true_iZ_i + \epsilon_i$, where $\epsilon_i \iid \Nor(0, \sigma^{\true, 2})$ with $\sigma^{\true, 2}=1$. 
In \eqref{eq:sim1-pi}, the parameter $t$ controls the strength of targeted selection. Targeted selection is a setting where $\pi^\true$ and $\mu^\true_i$ are related. We varied  the strength of targeted selection by changing the value of $t$, $t = -2, -1, 0, 1,$ and 2.  
Figure \ref{fig:Kang_mu_pi} plots $\pi_i^\true$ against $\mu^\true_i$ for each value of $t = -2, -1, 0, 1,$ and 2 using a simulated dataset. With a larger value of $t$, individuals with a higher expected outcome under no treatment are more likely to be assigned to treatment. When $t=0$, there is no direct functional relationship between $\pi^\true_i$ and $\mu^\true_i$. The case of $t=-1$ leads to the same setup as in \cite{kang2007demystifying}.
We designed the covariates $X_{ij}$, $j=1, \ldots, 4$, as those observable by a data analyst, to reflect realistic data analysis scenarios we commonly encounter; 
\begin{eqnarray*}
    &X_{i1} = \exp(W_{i1}/2), \hspace{0.75in} & X_{i2} = W_{i2}/\{1+\exp(W_{i1})\} + 10, \\
    &X_{i3} = (W_{i1}W_{i3}/25 + 0.6)^3, ~~ & X_{i4} = (W_{i2} + W_{i4} + 20)^2.
\end{eqnarray*}
We applied the four models including cLSBP to the dataset of $\{(X_i, Y_i), i=1, \ldots, N\}$. 

\begin{table}[!t]
  \centering
  \begin{tabular}{rll|ccc|ccc}
    \multicolumn{3}{r}{}
    & \multicolumn{3}{|c|}{ATE}
    & \multicolumn{3}{c}{CATE}
    \\ \toprule
    $t$ & Method & $\hat{\pi}$ & RMSE & COV & LEN & RMSE & COV & LEN
    \\ \midrule
    $t=-2$ & cLSBP & With & {\bf 0.40} & {\bf 0.71} & 0.92 & {\bf 1.10} & {\bf 0.87} & 3.36
    \\
    &  & Without & 0.40 & 0.70 & 0.92 & {\bf 1.10} & 0.86 & 3.32
    \\ \cmidrule{2-9}
    & BCF & With & 0.42 & 0.58 & 0.80 & 1.41 & 0.85 & 3.70
    \\
    & & Without & 0.54 & 0.32 & 0.78 & 1.47 & 0.83 & 3.67
    \\ \midrule
    $t=-1$ & cLSBP & With & 0.30 & 0.76 & 0.73 & {\bf 0.84} & {\bf 0.90} & 2.76
    \\
    &  & Without & 0.29 & 0.78 & 0.72 & 0.86 & 0.90 & 2.78
    \\ \cmidrule{2-9}
    & BCF & With & {\bf 0.25} & {\bf 0.82} & 0.68 & 1.28 & 0.86 & 3.48
    \\
    & & Without & 0.33 & 0.58 & 0.65 & 1.30 & 0.85 & 3.42
    \\ \midrule
    $t=0$ & cLSBP & With & {\bf 0.17} & 0.97 & 0.68 & {\bf 0.74} & {\bf 0.93} & 2.57
    \\
     & & Without & {\bf 0.17} & {\bf 0.96} & 0.67& 0.77 & 0.92 & 2.53
    \\ \cmidrule{2-9}
    & BCF & With & 0.19 & 0.92 & 0.66 & 1.24 & 0.86 & 3.37
    \\
    &  & Without & 0.19 & 0.88 & 0.62 & 1.23 & 0.87 & 3.35
    \\ \midrule
    $t=1$ & cLSBP & With & {\bf 0.15} & {\bf 0.97} & 0.70 & 0.80 & {\bf 0.92} & 2.63
    \\
     &  & Without & 0.16 & {\bf 0.97} & 0.69 & {\bf 0.79} & {\bf 0.92} & 2.59
    \\ \cmidrule{2-9}
    & BCF & With & 0.20 & {\bf 0.97} & 0.70 & 1.27 & 0.87 & 3.49
    \\
    &  & Without & 0.20 & 0.92 & 0.67 & 1.27 & 0.87 & 3.44
    \\ \midrule
    $t=2$ & cLSBP & With & {\bf 0.27} & {\bf 0.88} & 0.78 & {\bf 0.90} & {\bf 0.90} & 2.90
    \\
     &  & Without & 0.31 & 0.78 & 0.79 & 0.92 & 0.89 & 2.86
    \\ \cmidrule{2-9}
    & BCF & With & 0.35 & 0.74 & 0.79 & 1.35 & 0.86 & 3.64
    \\
    &  & Without & 0.40 & 0.60 & 0.77 & 1.36 & 0.85 & 3.56
    \\ \bottomrule
  \end{tabular}
  \caption{
    [Simulation 1] Root mean square estimation error (RMSE), coverage (COV), and average interval length (LEN) of the 95\% credible interval estimates are reported for both the average treatment effect (ATE) estimates and the conditional average treatment effect (CATE) estimates. The smallest RMSE and the COV closest to 95\% are marked in bold.
  }
  \label{table:Kang_result}
\end{table}

The results based on $R=100$ simulated datasets are summarized in Table~\ref{table:Kang_result}. They indicate that cLSBP outperformed BCF in the CATE estimation. cLSBP consistently achieved lower RMSE and shorter interval lengths, with better coverage across all values of $t$. For instance, at $t=2$, cLSBP obtained a lower RMSE (0.90 vs. 1.35) and a shorter interval length (2.90 vs. 3.64), with better coverage (0.90 vs. 0.86) compared to BCF. For the ATE estimation, cLSBP achieved a smaller RMSE except for the case where $t=-1$. For the cases with $t=1$ and $t=2$, cLSBP's improvement in ATE estimation relative to BCF was significant. Although BCF tended to have shorter intervals in some cases, this came at the cost of poorer coverage, especially in cases of strong targeted selection with $t=-2$ or $t=2$.
Inference for cases with larger values of $|t|$ can be more challenging due to stronger targeted selection, and cLSBP consistently showed superior performance compared to BCF in estimating both ATE and CATE for those cases.
Comparing cLSBP and BCF with and without $\hat{\pi}_i$ shows that incorporating $\hat{\pi}_i$ into the response estimation improves the ATE estimation for BCF when strong targeted selection is present ($t=-2$ or $t=2$). On the other hand, its effect on cLSBP's performance is minimal for the assumed scenarios. It also has little impact on the CATE estimation for both methods.

\subsection{Simulation 2}\label{sec:simulation2}

We considered the setup of the simulation studies in Section 6.1 of \cite{hahn2020bayesian}. We assumed five covariates, $X_{i1}, \ldots, X_{i5}$, for $i=1, \ldots, n$, where continuous covariates $X_{i1}$, $X_{i2}$, and $X_{i3}$ are generated from the standard normal distribution, a binary covariate $X_{i4}$ is generated from the Bernoulli distribution with a success probability of 0.5, and a trinary covariate $X_{i5}$ takes values 1, 2, or 3 with equal probability. We considered the following specifications for $\mu^{\true}_i$ and $\tau^{\true}_i$;
\begin{align*}
    \mu^\true_i &= \begin{cases}
    1+g(X_{i5})+X_{i1}X_{i3}, &\text{linear,} \\
    -6+g(X_{i5})+6 \mid X_{i3}-1 \mid, &\text{nonlinear,}
    \end{cases} \\
    \tau^\true_i &= \begin{cases}
    3, &\text{homogeneous,}\\
    1+2X_{i2}X_{i4}, &\text{heterogeneous,}
    \end{cases} 
\end{align*}
where $g$ is a function with values $g(1) = 2$, $g(2) = -1$, and $g(3) = -4$.
We then specified the true propensity score function, $\pi_i^\true = 0.8\Phi(3\mu^\true_i/s-0.5X_{i1})+0.05+u_i/10$, where $u_i \iid \Unif(0, 1)$ and $s$ is the standard deviation of $\mu_i^\true$'s taken over the simulated sample.
We finally generated $Z_i \mid \pi^\true_i \indep \Ber(\pi_i^\true)$ and $y_i = \mu^\true_i + \tau^\true_i Z_i + \epsilon_i$, where $\epsilon_i \iid \Nor(0, \sigma^{\true, 2})$ with $\sigma^{\true, 2}=0.25$. 
The two different specifications for both $\mu^\true_i$ and $\tau^\true_i$ result in four distinct data-generating processes. Also, we varied the sample size, $n=250$ and $500$, resulting in eight scenarios.

\begin{table}[!t]
  \centering
  \begin{tabular}{rrll|ccc|ccc}
    \multicolumn{4}{r}{}
    & \multicolumn{3}{|c|}{ATE}
    & \multicolumn{3}{c}{CATE}
    \\ \toprule
    $\tau^\true$ & $n$ & model & $\hat{\pi}$ & RMSE & COV & LEN & RMSE & COV & LEN
    \\ \midrule
    Homo-& $250$ & cLSBP & With & {\bf 0.14} & {\bf 0.98} & 0.65 & {\bf 0.26} & 0.99 & 1.42
    \\
    geneous && cLSBP & Without & {\bf 0.14} & 0.98 & 0.65 & {\bf 0.26} & {\bf 0.98} & 1.44
    \\ \cmidrule{3-10}
    && BCF & With & {\bf 0.14} & 0.90 & 0.50 & 0.36 & {\bf 0.98} & 1.28
    \\
    && BCF & Without & 0.16 & 0.84 & 0.47 & 0.36 & {\bf 0.98} & 1.21
    \\ \cmidrule{2-10}
    & $500$ & cLSBP & With & 0.11 & 0.93 & 0.40 & {\bf 0.21} & 0.98 & 1.01
    \\
    && cLSBP & Without & 0.11 & {\bf 0.94} & 0.39 & {\bf 0.21} & {\bf 0.97} & 1.01
    \\ \cmidrule{3-10}
    && BCF & With & {\bf 0.09} & 0.90 & 0.31 & 0.25 & 0.99 & 0.82
    \\
    && BCF & Without & 0.10 & 0.84 & 0.29 & 0.24 & 0.98 & 0.79
    \\ \cmidrule{1-10}
    Hetero- & $250$ & cLSBP & With & 0.21 & {\bf 0.94} & 0.77 & 0.62 & 0.97 & 2.24
    \\
    geneous && cLSBP & Without & 0.21 & {\bf 0.96} & 0.76 & {\bf 0.60} & 0.97 & 2.21
    \\ \cmidrule{3-10}
    && BCF & With & {\bf 0.16} & 0.90 & 0.56 & 0.62 & {\bf 0.94} & 2.02
    \\
    && BCF & Without & 0.19 & 0.81 & 0.52 & 0.63 & {\bf 0.94} & 1.94
    \\ \cmidrule{2-10}
    & $500$ & cLSBP & With & {\bf 0.09} & {\bf 0.97} & 0.44 & 0.49 & 0.97 & 1.62
    \\
    && cLSBP & Without & 0.10 & {\bf 0.97} & 0.44 & 0.50 & 0.96 & 1.58
    \\ \cmidrule{3-10}
    && BCF & With & {\bf 0.09} & 0.92 & 0.33 & {\bf 0.44} & {\bf 0.95} & 1.40
    \\
    && BCF & Without & 0.11 & 0.86 & 0.31 & {\bf 0.44} & 0.94 & 1.38
    \\ \bottomrule
  \end{tabular}
  \caption{[Simulation 2] A {\em linear} function is assumed for $\mu^\true$. Root mean square estimation error (RMSE), coverage (COV), and average interval length (LEN) of the 95\% credible interval estimates are reported for both the average treatment effect (ATE) estimates and the conditional average treatment effect (CATE) estimates. The smallest RMSE and the COV closest to 95\% are marked in bold.
  }
  \label{table:hahn_total_results_lin}
\end{table}

\begin{table}[!t]
  \centering
  \begin{tabular}{crll|ccc|ccc}
    \multicolumn{4}{r}{}
    & \multicolumn{3}{|c|}{ATE}
    & \multicolumn{3}{c}{CATE}
    \\ \toprule
    $\tau^\true$ & $n$ & Model & $\hat{\pi}$ & RMSE & COV & LEN & RMSE & COV & LEN
    \\ \midrule
    Homo- & $250$ & cLSBP & With & {\bf 0.08} & 1.00 & 0.67 & {\bf 0.12} & 1.00 & 1.18
    \\
    geneous&& cLSBP & Without & 0.09 & 1.00 & 0.66 & 0.25 & 1.00 & 1.19
    \\ \cmidrule{3-10}
    && BCF & With & 0.14 & {\bf 0.96} & 0.55 & 0.48 & {\bf 0.98} & 1.32
    \\
    && BCF & Without & 0.17 & 0.84 & 0.50 & 0.49 & {\bf 0.98} & 1.24
    \\ \cmidrule{2-10}
    & $500$ & cLSBP & With & {\bf 0.06} & 1.00 & 0.36 & {\bf 0.09} & 1.00 & 0.64
    \\
    && cLSBP & Without & {\bf 0.06} & 1.00 & 0.36 & 0.11 & 1.00 & 0.66
    \\ \cmidrule{3-10}
    && BCF & With & 0.09 & {\bf 0.93} & 0.32 & 0.35 & {\bf 0.98} & 0.82
    \\
    && BCF & Without & 0.09 & 0.88 & 0.30 & 0.32 & 0.99 & 0.77
    \\ \cmidrule{1-10}
    Hetero- & $250$ & cLSBP & With & 0.19 & 1.00 & 0.79 & 0.76 & 0.83 & 1.88
    \\
    geneous&& cLSBP & Without & 0.17 & 0.98 & 0.77 & {\bf 0.72} & 0.84 & 1.86
    \\ \cmidrule{3-10}
    && BCF & With & {\bf 0.16} & {\bf 0.97} & 0.61 & 0.80 & 0.93 & 2.15
    \\
    && BCF & Without & 0.21 & 0.82 & 0.55 & 0.79 & {\bf 0.94} & 2.06
    \\ \cmidrule{2-10}
    & $500$ & cLSBP & With & {\bf 0.09} & {\bf 0.96} & 0.43 & 0.53 & 0.93 & 1.16
    \\
    && cLSBP & Without & {\bf 0.09} & 0.98 & 0.42 & {\bf 0.51} & 0.94 & 1.14
    \\ \cmidrule{3-10}
    && BCF & With & {\bf 0.09} & {\bf 0.94} & 0.34 & 0.56 & 0.94 & 1.44
    \\
    && BCF & Without & 0.11 & 0.85 & 0.32 & 0.54 & {\bf 0.95} & 1.39
    \\ \bottomrule
  \end{tabular}
  \caption{
  [Simulation 2. Continued] A {\em nonlinear} function is assumed for $\mu^\true$. Root mean square estimation error (RMSE), coverage (COV), and average interval length (LEN) of the 95\% credible interval estimates are reported for both the average treatment effect (ATE) estimates and the conditional average treatment effect (CATE) estimates. The smallest RMSE and the COV closest to 95\% are marked in bold.
  }
  \label{table:hahn_total_results_non}
\end{table}

The results, based on 200 simulated datasets for each scenario, are presented in Tables \ref{table:hahn_total_results_lin} and \ref{table:hahn_total_results_non} for the linear and nonlinear $\mu_i^\true$ specifications, respectively. 
\cite{hahn2020bayesian} does not provide a detailed hyperparameter specification of BCF used for the simulation study, and the default specification in the R package is used for the simulation study of this section. Possibly due to different specifications of the tree models, the results in the tables differ from those in Tables 2 and 3 of \cite{hahn2020bayesian}.
Overall, the results indicate that cLSBP has a favorable performance compared to BCF. When it does not achieve better performance, its performance is very close to that of BCF in most cases.
Coverage for both models approaches the desired levels as the sample size increases, although cLSBP consistently delivers shorter credible intervals while maintaining adequate coverage.
For all methods, performance improves with a larger sample size, and the improvement is more pronounced in the heterogeneous cases.
It is also notable that incorporating $\hat{\pi}_i$ improves the performance of BCF, especially for the estimation of ATE when $n$ is small, but does not have a substantial effect on the performance of cLSBP in this simulation study.

\section{Data Analyses} \label{sec:data-analyses}
\subsection{Reanalysis of Data from the 2022 ACIC Data Challenge}\label{sec:ACIC}
We applied cLSBP to the datasets simulated for the 2022 American Causal Inference Conference (ACIC) Data Challenge. The challenge is designed to evaluate causal impacts using simulated datasets that resemble real-world data from evaluations of large-scale U.S. healthcare interventions aimed at reducing Medicare expenditures. These datasets include four years of beneficiary-level Medicare expenditure data, with beneficiaries clustered in approximately 500 primary care practices. They incorporated both patient- and practice-level covariates, with intervention participation determined at the practice level using combined confounders after two years. The
intervention is assigned at the practice level and takes effect only in the post-intervention period(Years 3 and 4).

A range of data generating processes were considered by varying the sources and magnitudes of confounding and heterogeneity, including a scenario similar to a randomized trial, resulting in 17 different data generating processes. For each DGP, 200 datasets were simulated, providing a total of 3,400 datasets to challenge participants. Specifically, participants were required to estimate three estimands: (i) the SATT, $\tau_0=1/M\sum_{t=3}^4\sum_{j:z_j=1}Y_{jt}(1)-Y_{jt}(0)$, (ii) the subgroup CATTs, $\tau_s=1/M_s\sum_{t=3}^4\sum_{j\in s}Y_{jt}(1)-Y_{jt}(0)$ for each of the 14 prespecified practice-level subgroups, and (iii) the practice-level UTE, $\tau_{i}=1/M_i\sum_{t=3}^4\sum_{j:z_j=1,j\in i} Y_{jt}(1)-Y_{jt}(0)$, where $i$ indexes practices, $j$ indexes patients, $t\in\{3,4\}$ denotes the post-treatment years, and $M, M_s$, and $M_i$ are the corresponding patient totals over years 3-4. Methods were evaluated based on various criteria, including RMSE and coverage of interval estimates. The datasets, along with their ground truths, are publicly available. \cite{thal2023causal} provides a detailed description of the data generating processes and reports that methods employing tree-based approaches performed best, and model-based interval estimators generally outperform bootstrap-based ones. \cite{kokandakar2023bayesian} modified BCF and reanalyzed the challenge datasets. They explored how BCF's performance changes by different modeling choices, such as the estimation method of the propensity score and the use of sparsity-inducing priors.

We reanalyzed the 3,400 datasets aggregated at the practice level. Following the competition's instructions, the practice-level estimates were weighted by the number of patients in each practice to obtain an estimate of patient-level impacts. We also performed propensity sensitivity analyses to assess the performance based on the choice of propensity score estimation method. Similar to those in \cite{kokandakar2023bayesian}, we employed BART, Covariate Balancing Propensity Score (CBPS, \cite{imai2014covariate}), Gradient Boosting Machine (GBM, \cite{gbm2024package}), and LASSO (\cite{friedman2010regularization}) for the propensity score estimation, referring to each as cLSBP-BART, cLSBP-CBPS, cLSBP-GBM, and cLSBP-LASSO, respectively. In addition, we included cLSBP without $\hat{\pi}$, referred to as cLSBP-no. 

\begin{figure}[!t]
    \centering
    \begin{tabular}{c}
    \includegraphics[width=0.95\linewidth]{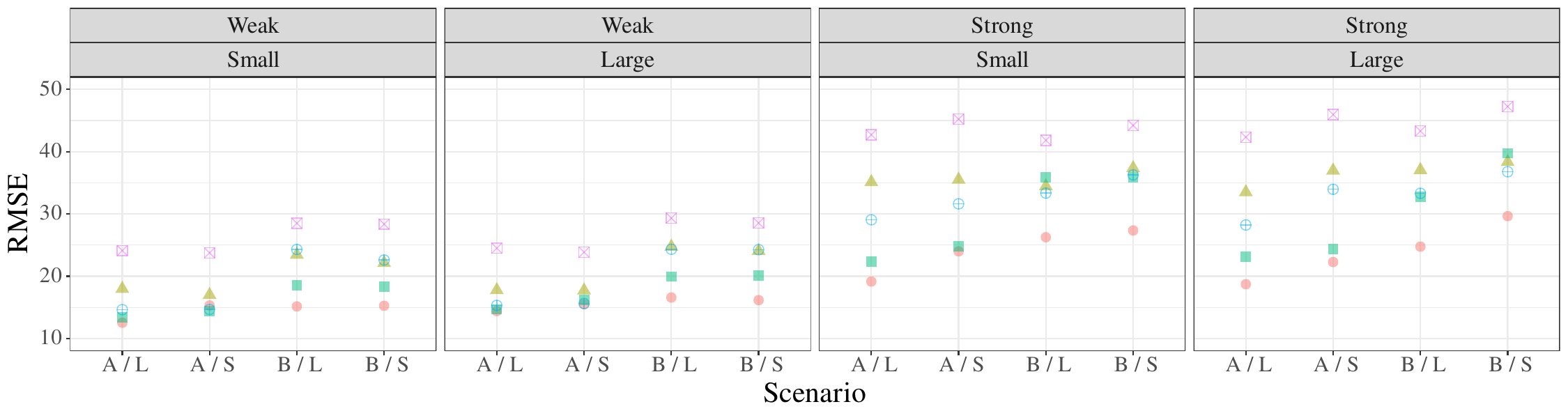}\\
    (a) RMSE\\
    \includegraphics[width=0.95\linewidth]{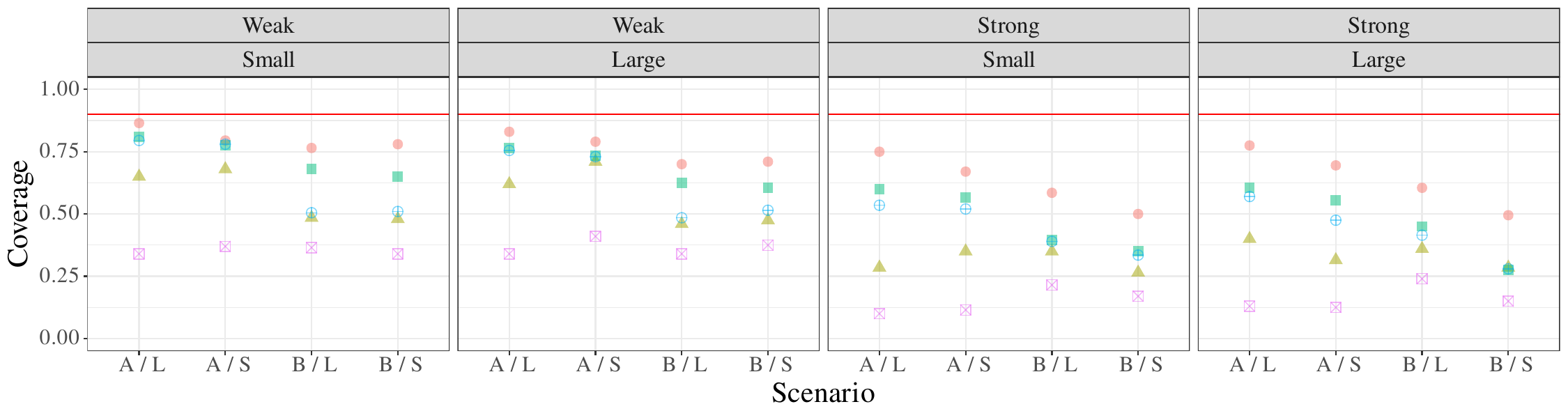}\\
    (b) Coverage \\
    \includegraphics[width=0.95\linewidth]{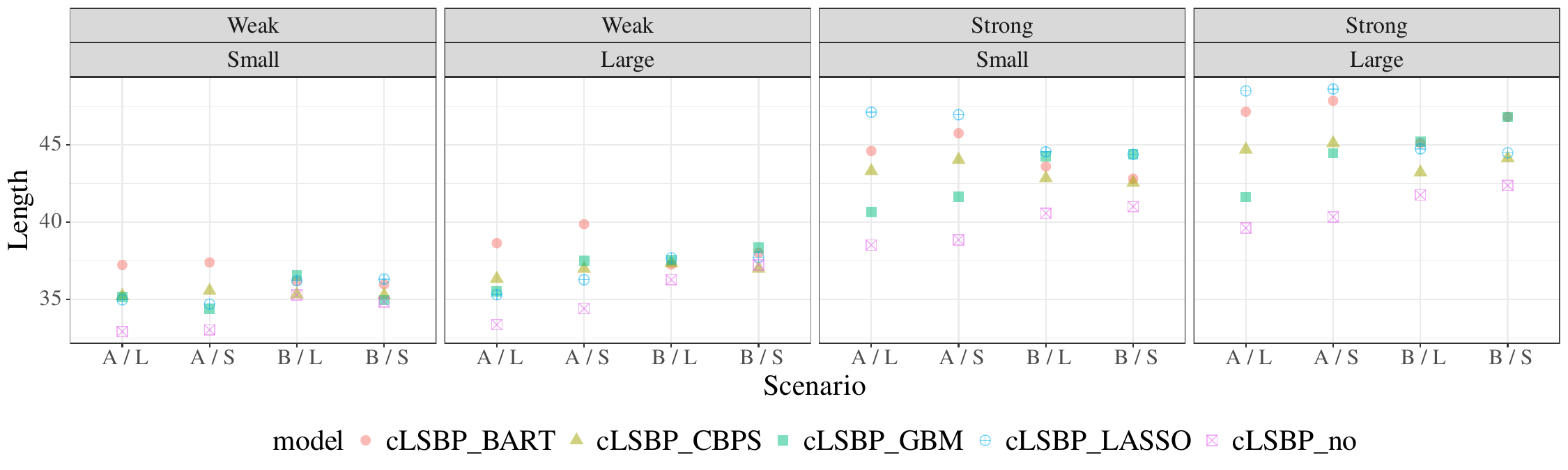}\\
    (c) Length
    \end{tabular}
    \caption{[The 2022 ACIC Data Challenge] The RMSE of SATT estimates, and the coverage and length of their 90\% interval estimates, are plotted across the 16 confounded scenarios in panels (a), (b), and (c), respectively. The four subpanels in each panel represent different combinations of confounding magnitude and heterogeneity: Strong vs.\ Weak confounding and Small vs.\ Large heterogeneity. Each subpanel includes four scenarios based on the sources of confounding and heterogeneity, where the two different sources are denoted by A and B for confounding and L and S for heterogeneity. cLSBP is applied with four different choices, BART, CBPS, GBM, and LASSO for the estimation method of $\pi$, and cLSBP without propensity score estimates is included for comparison.}
    \label{fig:ACIC_overall}
\end{figure}

\begin{figure}[!t]
  \centering
\begin{tabular}{cc}
   \includegraphics[width=0.45\linewidth]{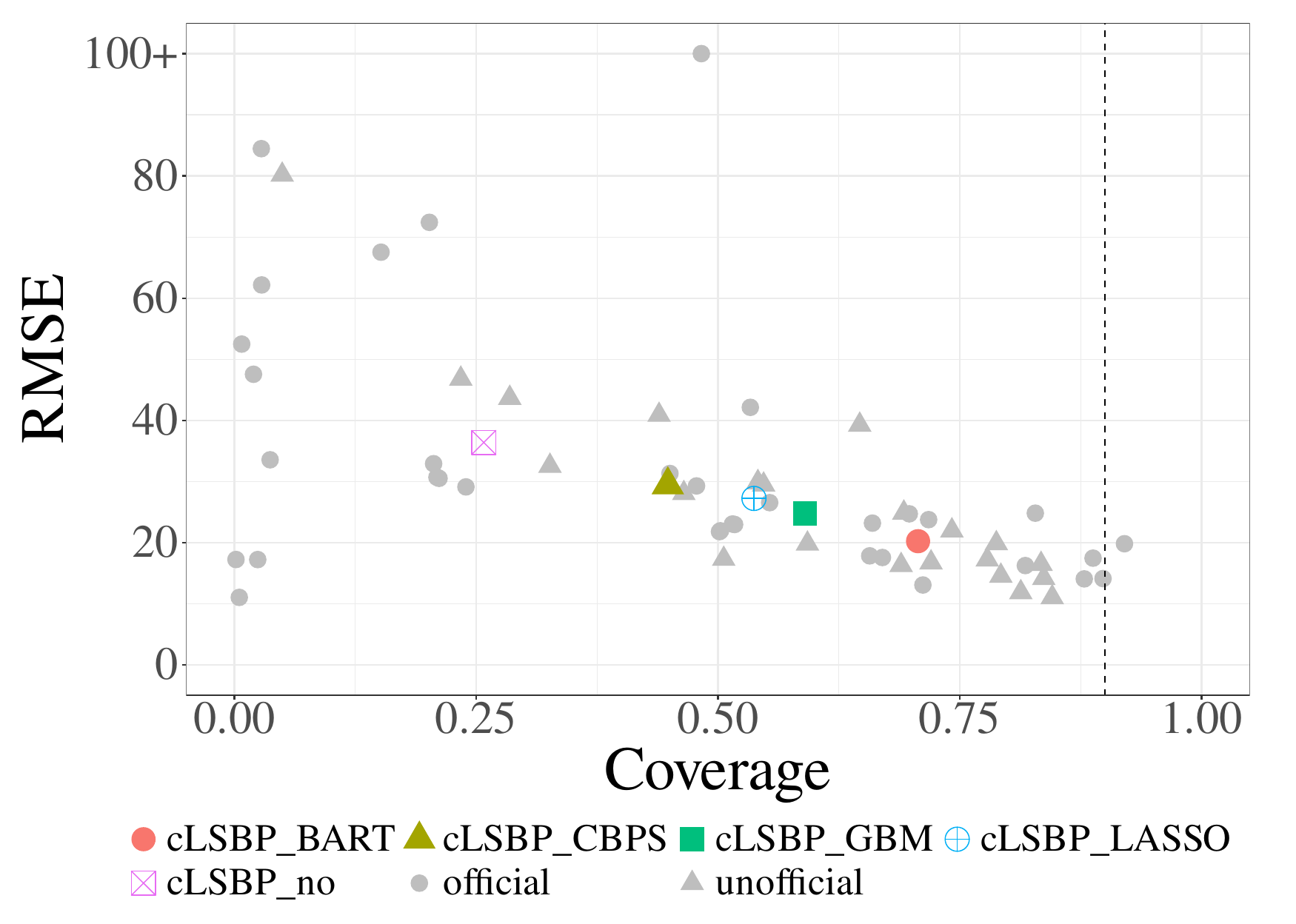}&
   \includegraphics[width=0.45\linewidth]{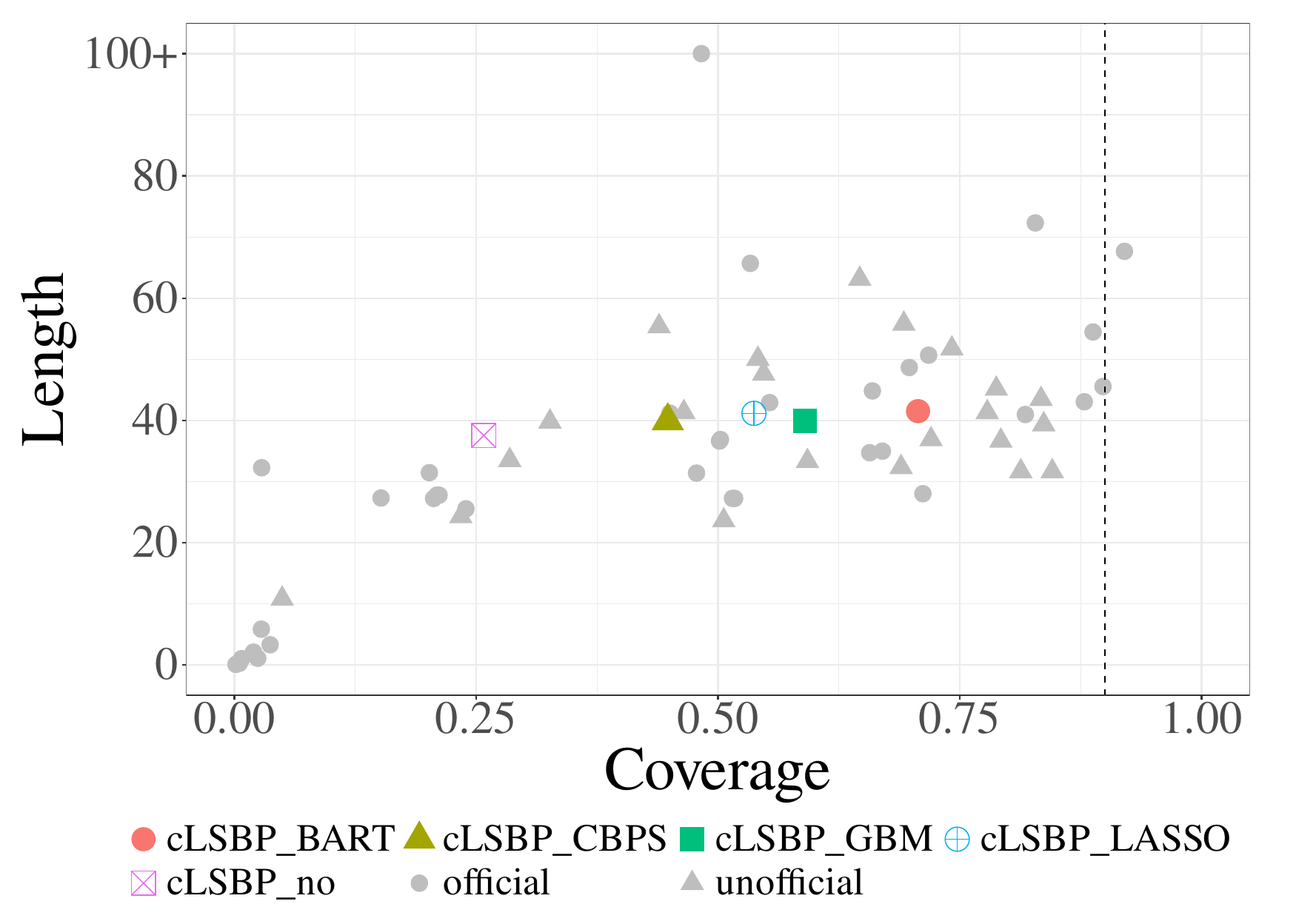}\\
       (a) COV vs RMSE & (b) COV vs LEN   \\
  \end{tabular}
 \caption{[The 2022 ACIC Data Challenge] Estimates of SATT under the cLSBP-based models are compared with those under the challenge participating methods under the criteria of RMSE, coverage (COV), and length (LEN). The dashed vertical lines represent the nominal coverage of 90\%. 
 } 
\label{fig:ACIC-comp}
\end{figure}

\begin{figure}[!t]
    \centering
    \includegraphics[width=\linewidth]{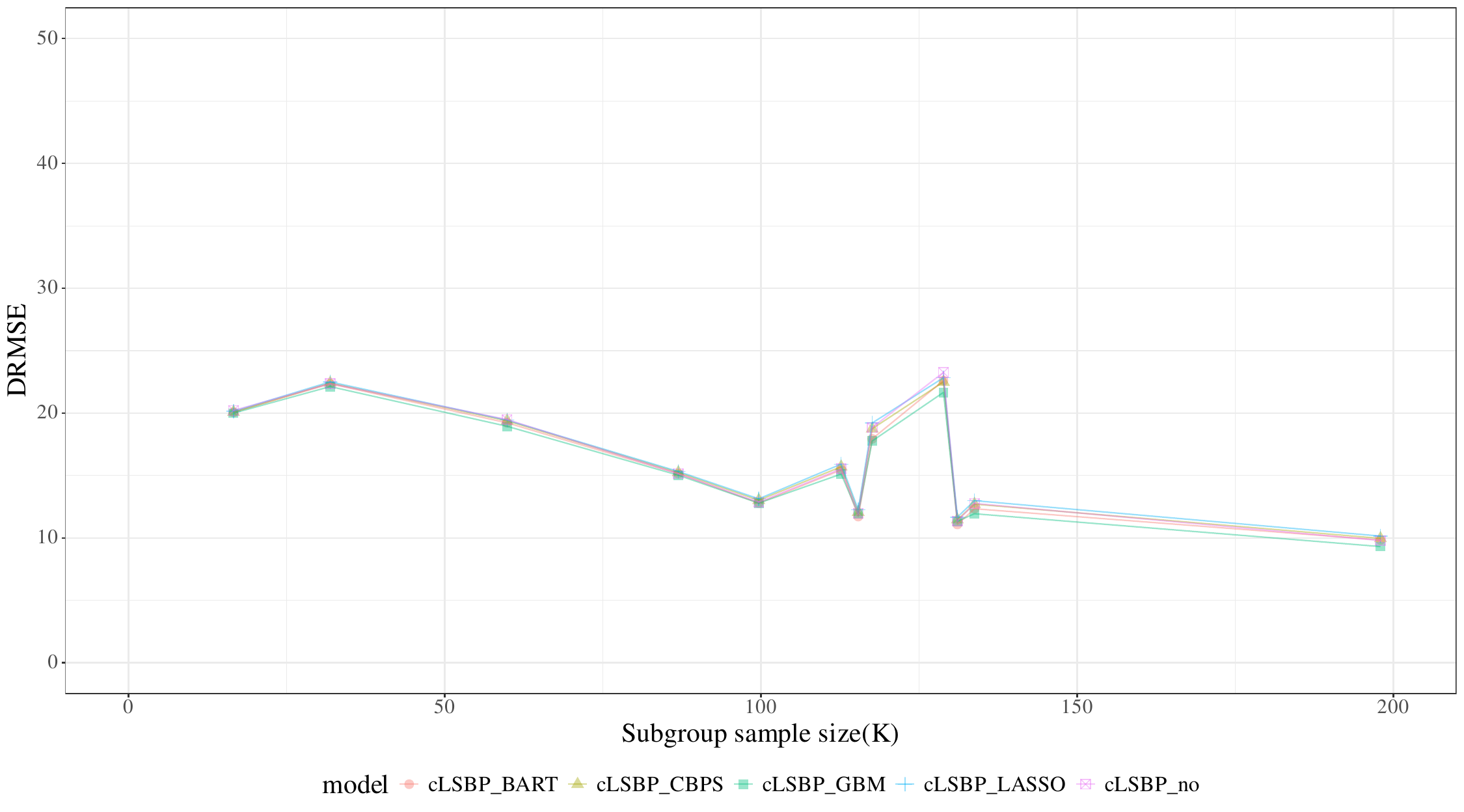}
    \caption{[The 2022 ACIC Data Challenge]  The differential root mean squared errors (DRMSEs) for each of the prespecified practice subgroups are plotted over subgroup sizes averaged over 3,400 datasets.  The cLSBP-based models with four different $\pi$ estimation methods, as well as the model without $\hat{\pi}$, are used. }
    \label{fig:subgroup_CATT_DRMSE}
\end{figure}

\begin{figure}[!t]
  \begin{center}
\begin{tabular}{cc}
    \includegraphics[width=0.45\linewidth]{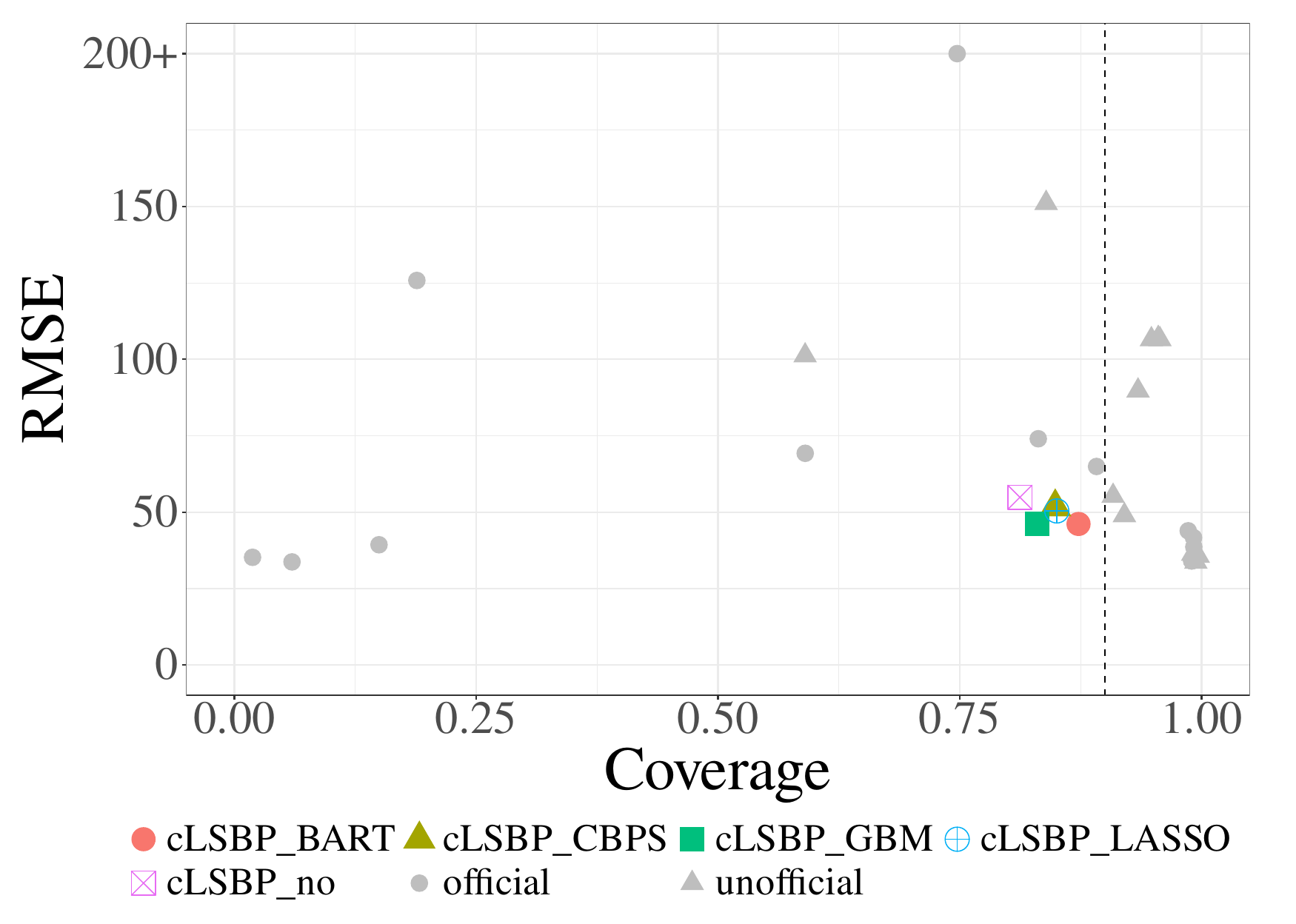} &
    \includegraphics[width=0.45\linewidth]{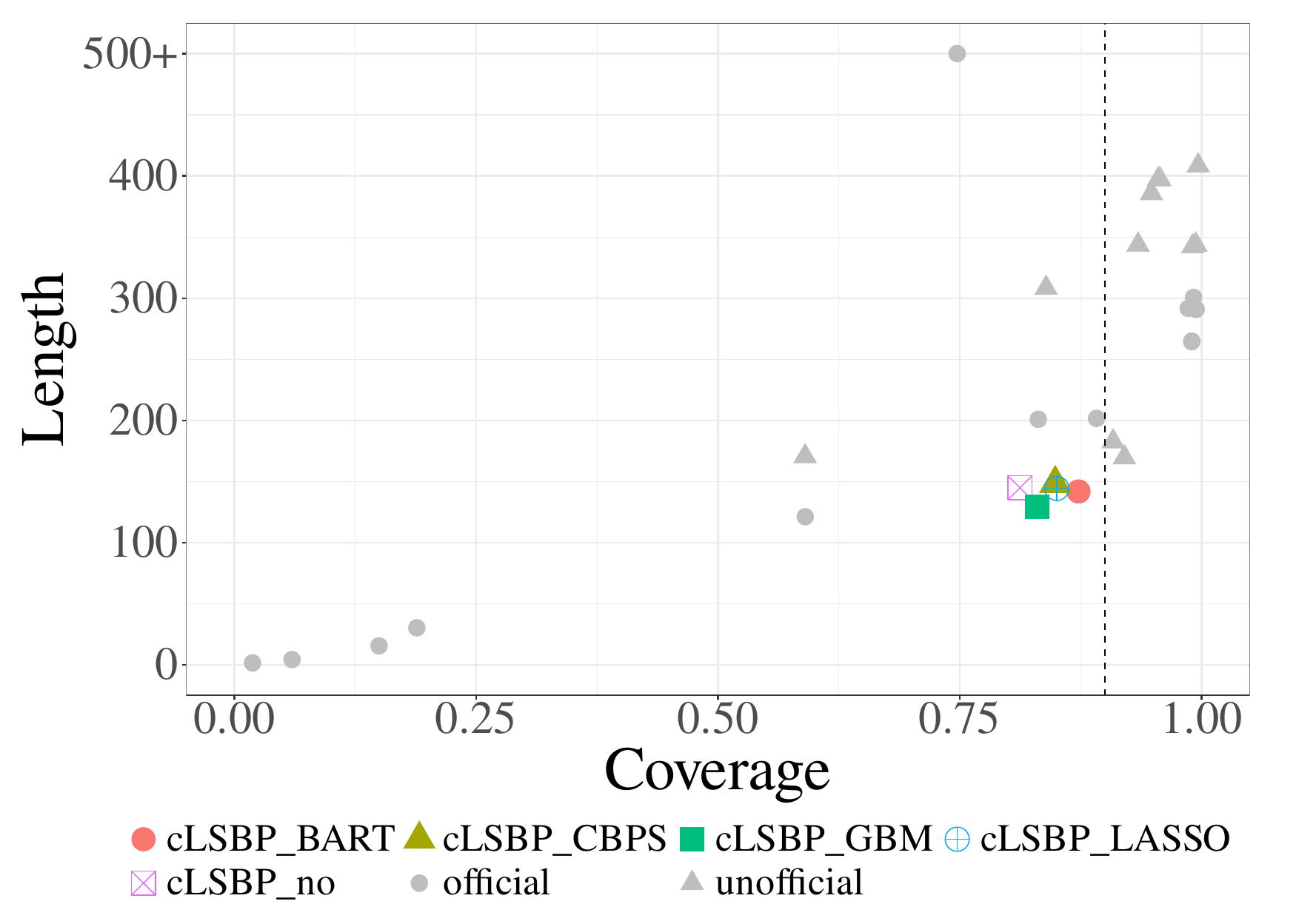} \\
    (a) COV vs RMSE & (b) COV vs LEN \\
  \end{tabular}
 \end{center}
 \vspace{-0.15in}
 \caption{[The 2022 ACIC Data Challenge] Estimates of individual unit treatment effects (UTEs) at the practice level under the cLSBP-based models are compared with those under the challenge participating methods under the criteria of RMSE, coverage (COV), and length (LEN). The dashed vertical lines represent the nominal coverage of 90\%.} 
\label{fig:UTE}
\end{figure}

Figure \ref{fig:ACIC_overall} (a), (b) and (c) illustrate RMSE of overall SATT estimates, and COV and LEN of their 90\% interval estimates under the five different cLSBP models for the 16 confounded scenarios, respectively. The four subpanels in each of the panels correspond to different combinations of the magnitude of confounding and heterogeneity: `Strong' and `Weak' for confounding and `Small' and `Large' for heterogeneity. Each subpanel presents the results for the four different scenarios based on the sources of confounding and heterogeneity, where the two sources are denoted by `A' and `B' for confounding and `L' and `S' for heterogeneity. The figures show that incorporating the propensity score estimate greatly improves performance. More interestingly, the performance of cLSBP varies significantly with the choice of estimation method for $\pi$. cLSBP-BART performs best in all 16 scenarios for RMSE and coverage. 

In addition, Figure \ref{fig:ACIC-comp} compares the performance of cLSBP with that of the 58 estimators that participated in the competition. These estimators are based on a wide range of approaches, including machine learning methods, semiparametric and nonparametric models for the response surface, as well as methods focused on treatment assignment modeling or doubly robust estimation. The comparison is based on RMSE, COV, and LEN of the SATT estimates. The figure shows that cLSBP with $\hat{\pi}$ performs competitively overall.

We next examined the performance of cLSBP for CATTs in the pre-specified practice subgroups by computing the differential root mean squared error (DRMSE), as suggested in \cite{thal2023causal}; $\mbox{DRMSE}_s =  \sqrt{\mbox{E}[\{(\widehat{\tau_s} - \widehat{\tau}_0) - (\tau_s-\tau_0)\}^2]}$, for subgroup $s$, where $\tau_0$ is the true overall SATT, $\tau_s$ is the true CATT for subgroup $s$, and $\widehat{\tau}_0$ and $\widehat{\tau}_s$ are their respective estimates. This metric focuses on estimating heterogeneity among subgroups. As shown in Figure~\ref{fig:subgroup_CATT_DRMSE}, the DRMSE for the cLSBP-based methods does not vary much across datasets regardless of sample sizes, indicating that they produce reliable subgroup-specific causal estimates, irrespective of subgroup size. Furthermore, when compared to the DRMSEs under the other methods shown in Figure 2 of \cite{thal2023causal}, the performance of the cLSBP-based models is very competitive and comparable to top-performing methods or benchmark methods such as gumpy and wbcf.

Lastly, we compared the individual unit treatment effect (UTE) estimates for treated practices from the cLSBP-based methods to those from the participating methods. In our implementation, we estimate the UTE estimand $\tau_i$ by averaging these practice-level posterior CATEs across Years 3 and 4 without reweighting by the year-specific patient counts, $\widehat{\tau}_i=\tfrac{1}{2}\bigl(\tau_{i,3} + \tau_{i,4})$, where $\tau_{i,t}=1/M_{it}\sum_{j\in i}Y_{jt}(1)-Y_{jt}(0)$ is the year-specific practice-level treatment effect. 

We report RMSE, coverage, and interval length relative to the official ACIC UTE truth. Similar to the challenge in estimating CATT for small subgroups, where overfitting is a concern, estimating UTEs is also difficult. Only 25 out of 58 estimators were submitted due to this difficulty, and shrinkage estimators among them performed better. Figure~\ref{fig:UTE} shows that the cLSBP-based methods dominate the participating methods for the UTE estimation in all criteria: RMSE, COV, and LEN. They achieved the smallest RMSE while maintaining coverage close to the nominal level, with the shortest interval length. Notably, the choice of a $\pi$ estimation method had little impact on performance in the estimation of UTEs, and cLSBP without $\hat{\pi}$ performed comparably to cLSBP with $\hat{\pi}$.

\subsection{Effect Of 401(k) Retirement Plans on Asset Accumulation}\label{sec:pension}

We applied our method to the Pension 401(k) dataset, available in the R package \textit{hdm}. The dataset is from the 1991 Survey of Income and Program Participation (SIPP) and is used to examine the effects of 401(k) plans on wealth. 401(k) plans are among the most popular tax-deferred programs. Since they are offered by employers, only workers at firms that provide such programs are eligible. The dataset has been analyzed using various causal inference methods, primarily with instrumental variable (IV) estimators that use 401(k) eligibility as an instrumental variable to estimate the effects of 401(k) retirement programs on savings (see \citet{abadie2003semiparametric}, \citet{belloni2017program}, \citet{sant2022covariate}, \citet{wang2024semiparametric} among many others for examples). In contrast, we considered only households whose reference persons are eligible for the 401(k) plan and focused on estimating the effect of participation in 401(k) plans on net financial assets ($y$), accounting for potential confounders that influence the decision to enroll in a 401(k). We let $z=1$ denote the treatment, where a household participates in a 401(k), and $z=0$ otherwise. The potential confounders ($x$) include age ($x_1$), income ($x_2$), family size ($x_3$), education ($x_4$), and binary indicators for the following: defined benefit pension participation status ($x_5$), marital status ($x_6$), two-earner status ($x_7$), participation in Individual Retirement Accounts ($x_8$), and homeownership ($x_9$). The covariates of age, income, and education are categorized, with age having 5 levels, income 7 levels, and education 3 levels. Before fitting the model to the dataset, we removed observations with extremely small or large values of net financial assets; specifically, we discarded observations where the absolute distance from the average net financial assets was in the top 5\%. $\hat{\pi}_i$ was estimated by BART. 

\begin{table}[!t]
\centering
\begin{tabular}{rll|ll}
\hline
 & \multicolumn{2}{c|}{cLSBP} & \multicolumn{2}{c}{BCF} \\
\toprule
 ATE  & 12.1 & (10.3, 13.7) & 12.8 & (11.1, 14.4) \\
\midrule
\multicolumn{5}{c}{(a) ATE} \\ \hspace{0.1in}
\end{tabular}

{\small 
\begin{tabular}{rlrll|ll}
\hline
Subgroup & & Size (\%) & \multicolumn{2}{c|}{cLSBP} & \multicolumn{2}{c}{BCF} \\
\cmidrule(r){4-5} \cmidrule(l){6-7}
& & & CATE & 95\% CrI & CATE & 95\% CrI \\
\toprule
Age & 25-29 & 395 (11\%) & 6.7 & (5.4, 8.0) & 6.8 & (2.5, 12.1) \\
& 30-34 & 725 (21\%) & 9.2 & (7.7, 10.5) & 7.7 & (4.3, 11.2) \\
& 35-44 & 1216 (35\%) & 12.4 & (10.5, 14.2) & 14.5 & (11.5, 16.9) \\
& 45-54 & 792 (23\%) & 14.7 & (12.2, 17) & 16.3 & (13.7, 19.4) \\
& 55-64 & 369 (11\%) & 16.9 & (13.7, 19.7) & 15.9 & (13.3, 19.3) \\
\midrule
Income & 0-10k & 43 (1\%) & 6.1 & (3.8, 9.1) & 7.8 & (-0.2, 15.6) \\
& 10-20k & 355 (10\%) & 5.4 & (4.4, 6.4) & 6.3 & (1.9, 10.5) \\
& 20-30k & 627 (18\%) & 7.0 & (5.9, 8.1) & 8.1 & (4.8, 11.4) \\  
& 30-40k & 706 (20\%) & 10.5 & (9, 11.9) & 11.5 & (8.2, 14.7) \\
& 40-50k & 562 (16\%) & 12.6 & (10.7, 14.4) & 12.3 & (8.7, 15.9) \\
& 50-75k & 835 (24\%) & 16.5 & (13.8, 19.1) & 17.3 & (13.9, 20.6) \\
& 75k+ & 369 (11\%) & 19.9 & (16, 23.5) & 20.5 & (16.3, 25.9) \\ 
\midrule
Education & No High School & 254 (7\%) & 9.5 & (7.8, 11.2) & 11.1 & (8.7, 15.4) \\ 
& High School & 1261 (36\%) & 10.7 & (9.0, 12.2) & 11.8 & (9.8, 13.7) \\  
& Some College & 907 (26\%) & 11.6 & (9.9, 13.2) & 12.4 & (10.5, 14.5) \\
& College & 1075 (31\%) & 14.7 & (12.4, 16.8) & 14.5 & (12.5, 16.6)\\ 
\bottomrule 
\multicolumn{7}{c}{(b) CATE} \\ 
\end{tabular}
}
\caption{[401(K) Dataset] Estimates of the average treatment effect (ATE) are shown in subtable (a). In subtable (b), the estimates of the average treatment effects across various subgroups (CATE) are presented. The subgroups are defined based on categorical variables such as age, income level, and education level. Their point estimates and 95\% credible interval estimates are obtained using cLSBP and BCP. The estimates are reported in units of 1000 USD.}
\label{tab:pension_cate_subgroups}
\end{table}

Table \ref{tab:pension_cate_subgroups} summarizes inferences on the effects of 401(k) retirement programs ($z$) on net financial assets ($y$). Subtable (a) presents the point estimate of ATE along with its 95\% CrI, indicating that participation in 401(k) plans increases net financial assets. Subtable (b) illustrates the CATE estimates for subgroups defined by age, income, and education. The effect remains positive for all subgroups, but its size varies across them. In general, as the age group increases or as income and/or education levels rise, the effect of participation is greater. The length of the 95\% CrI estimates remains fairly consistent across subgroups.

For comparison, we applied both cLSBP and BCF methods to the dataset, and their estimates are included in Table \ref{tab:pension_cate_subgroups}. The ATE estimates are similar under both methods. The CATE point estimates also align for large subgroups. However, for smaller subgroups, the differences in estimates are more pronounced.  The point estimates under the two methods are more different, and the CrIs are notably wider under BCF. For example, for the income subgroup of 0-10k, the point estimates are 6.1 and 7.8 under cLSBP and BCF, respectively, with 95\% CrI estimates of (3.8, 9.1) and (-0.2, 15.6). Even when the point estimates are similar, e.g., 6.7 vs 6.8 for the subgroup of age 25-29, their CrI estimates are vastly different, (5.4, 8.0) vs (2.5, 12.1).

\begin{figure}[!t]
    \centering
    \begin{tabular}{cc}
          \includegraphics[width=0.45\linewidth]{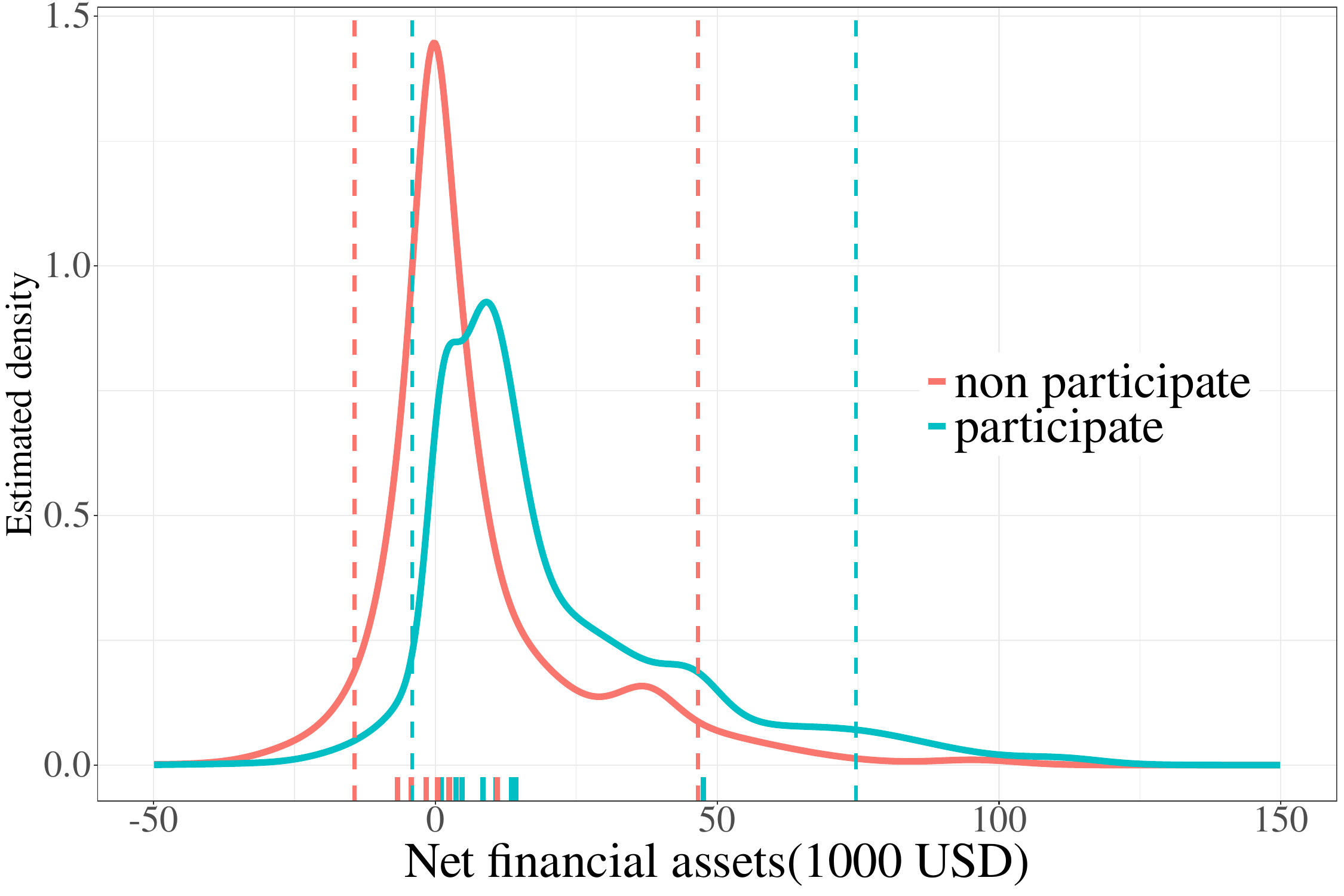}   & 
              \includegraphics[width=0.45\linewidth]{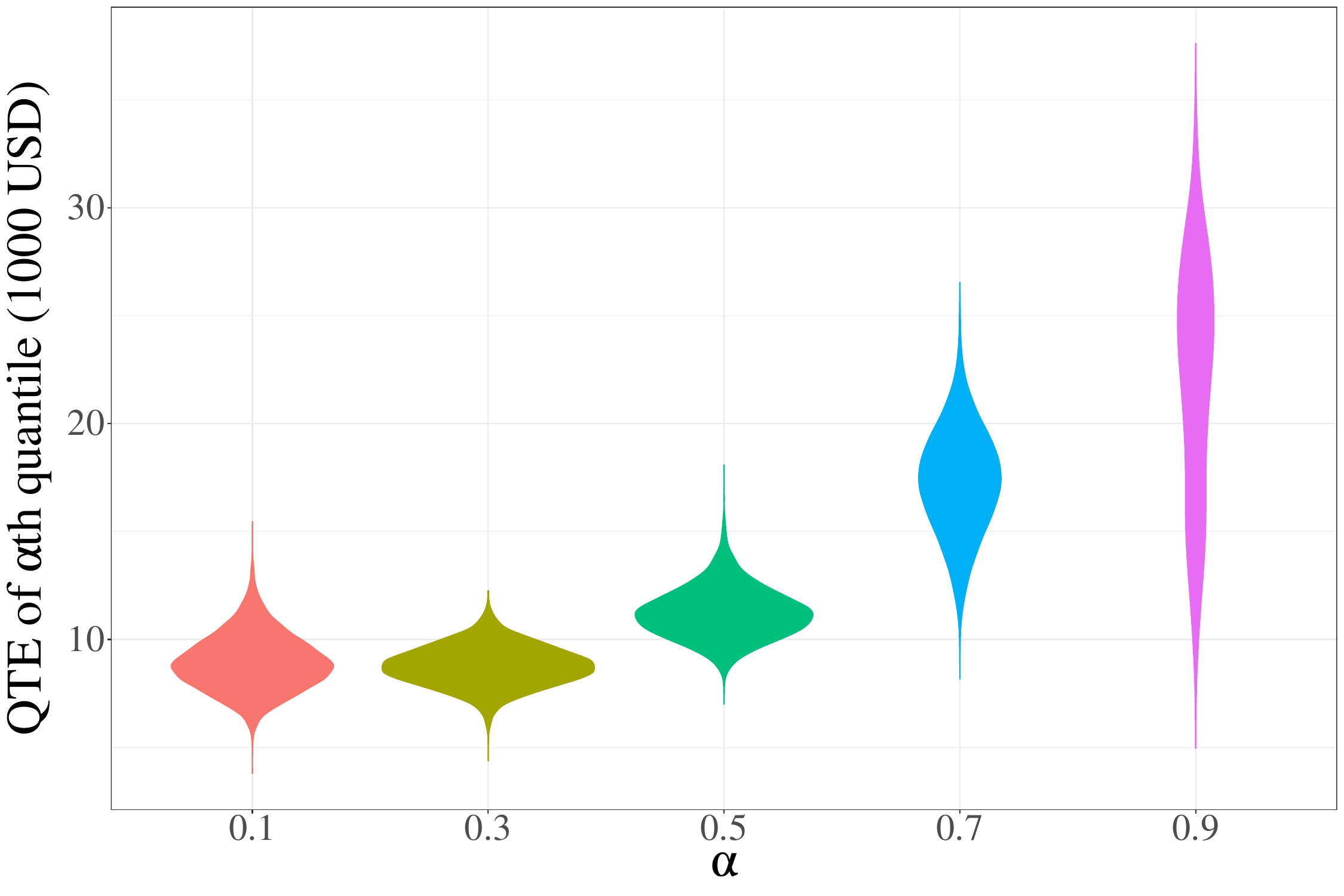}\\
         (a) Posterior Predictive Distribution & 
         (b) Quantile Treatment Effects \\ 
    \end{tabular}
    \caption{[401(K) Dataset] Posterior predictive distribution estimates of net financial assets for a household with a particular $x$ value ($x^\star$) are shown in panel (a). The observed net financial assets for $x=x^\star$ are displayed as a rug plot on the x-axis. The colors red and blue denote the control and treatment groups, respectively. \textit{Dashed vertical lines indicate the 90\% credible intervals of the posterior predictive distributions for each group.} The posterior distributions of quantile treatment effects (QTEs) at the 0.1, 0.3, 0.5, 0.7, and 0.9th quantiles, controlling for $x = x^\star$, are shown in panel (b).}
    \label{fig:pension_ppy_16subjects}
\end{figure}

In addition, we performed posterior predictive checking to assess the model fit. Specifically, we considered $x^\star$, a particular value of the confounders, and examined the posterior predictive distribution of $y$ for each of $z \in \{0, 1\}$, while controlling for the given $x^\star$. Estimates of the posterior predictive distributions are shown in Figure~\ref{fig:pension_ppy_16subjects}(a). The colors red and blue denote the control and treatment groups, respectively. We fixed $x^\star$ to represent a specific subjects with the following covariate values, ordered by $x_i$ for $i=1,\ldots,9$: aged 35–44 ($x_1$), income between 20--30k ($x_2$), family size 4 ($x_3$), high school education ($x_4$), not participating in a defined benefit pension ($x_5$), married ($x_6$), in a two-earner household ($x_7$), not participating in Individual Retirement Accounts ($x_8$), and a homeowner ($x_9$). Also, as an additional covariate, the estimated propensity score at $x^\star$ is $\hat{\pi}(x^\star) = 0.7417$. The observed net financial assets values with $x = x^\star$ are plotted at the bottom in a rug plot; there are 16 such observations in total, with 9 in the treatment group ($z=1$) and 7 in the control group ($z=0$). Despite the subjects having the same values of $x$, the observed net financial assets exhibit substantial variability within each group, which is reflected in the predictive distribution estimates. While participation has a positive effect on net financial asset values, this effect is not straightforward. The entire distribution—including its center and dispersion—changes with participation status. This illustration supports the need for a flexible model, such as cLSBP, to effectively capture non-standard density shapes. It strongly suggests that any strong distributional assumption is not suitable for analyzing this dataset.

A BNP approach enables principled inference on predictive distributions, allowing for a comprehensive exploration of treatment effects across the outcome distribution with uncertainty quantification. To illustrate this, we further examined quantile treatment effects (QTEs), $QTE(\alpha \mid x) = q_{Y(1)}(\alpha \mid x) - q_{Y(0)}(\alpha \mid x)$, for the 401(k) participation of a household with $x^\star$, where the conditional quantile function $q_{Y(z)}(\alpha \mid x)$ is the $\alpha$-th quantile of the distribution of $Y$ given $x$ and $z$. Figure~\ref{fig:pension_ppy_16subjects}(b) illustrates the posterior distribution of $QTE(\alpha \mid x^\star)$ for $\alpha \in \{0.1, 0.3, 0.5, 0.7, 0.9\}$. From the plots, participation results in an increase in net financial assets across all $\alpha$ values, and the effects are greater for larger values of $\alpha$.

\begin{table}[!t]
\centering
\begin{tabular}{lccc}
\hline
\textbf{Model} & \textbf{RMSE (sd)} & \textbf{Coverage (sd)} & \textbf{Length (sd)} \\ 
\hline
cLSBP & 23.2 (0.886) & 0.945 (0.008) & 84.3 (28.30) \\ 
BCF   & 22.6 (0.984) & 0.376 (0.020) & 14.7 (2.57) \\ 
\hline
\end{tabular}
\caption{[401(K) Dataset] The 5-fold cross-validation results for the BCF and cLSBP models are presented. The RMSE, coverage, and average length of interval estimates are shown, with standard deviations in parentheses.}
\label{tab:cv_results}
\end{table}

We further performed 5-fold cross-validation using cLSBP and BCF to compare their predictive performance on the dataset. The results are shown in Table \ref{tab:cv_results}. While RMSE is similar for both models, their coverage and the length of the interval estimates differ significantly. In particular, cLSBP provides better uncertainty quantification and achieves the nominal coverage.

\section{Conclusion}
In this paper, we proposed a flexible BNP approach for estimating various causal effect quantities, including ATE and CATE. Through simulation studies and a reanalysis of the 2022 ACIC Data Challenge data, the proposed cLSBP model demonstrated strong and consistent performance in estimating causal effects, especially HTE. In particular, cLSBP showed effective at the most granular level, such as practice-level unit treatment effects, where estimation is especially challenging due to small subgroup sizes. Furthermore, the analysis of 401(k) retirement plan data illustrates how cLSBP can be used to infer a variety of causal effect quantities.

The cLSBP framework also lends itself to several promising extensions. First, it can be readily applied to settings with non-binary treatments—such as categorical, ordinal, or continuous exposures—by appropriately modifying the treatment indicator while preserving the underlying mixture structure. Second, the current model adopts a relatively simple linear specification for the mixture atoms, with global shrinkage induced via a horseshoe prior. This can be replaced by more expressive priors, such as Gaussian process (GP) priors, to allow for local nonlinearity and higher-order interactions. These extensions remain compatible with Pólya-Gamma data augmentation, preserving the computational tractability of posterior inference while retaining the interpretability of covariate effects through the sequential logistic structure of LSBP. Taken together, these directions suggest that cLSBP is not only practically robust in its current form but also provides a versatile foundation for modeling increasingly complex causal structures in observational data.

\appendix

\makeatletter
\renewcommand{\thetheorem}{\Alph{section}.\arabic{theorem}} 
\makeatother




\section{Derivation of the Full Conditionals in Result \ref{propposterior}}
\label{appendixa}
In this section, we summarize the definition and property of the P\'olya-Gamma distribution, the prior distribution of $\omega_{i,h}$ in \cite{polson2013bayesian} and provide a derivation of Result \ref{propposterior}  in Section \ref{sectionposterior}.
\begin{definition}[P\'olya-Gamma Distribution]
        Given $b > 0$, let $g_n \overset{iid}{\sim} \mathrm{Gamma}(b, 1)$ for $n \in \mathbb{N}$. Then,
        \begin{align*}
            X &:= \frac{1}{2\pi^2}\sum_{n=1}^\infty \frac{g_n}{(n-1/2)^2+c^2/(4\pi^2)}
        \end{align*}
        follows P\'olya-Gamma distribution with parameters $b, c$ where $c \in \mathbb{R}$. In other words, $X \sim \mathrm{PG}(b, c)$.
\end{definition}

\begin{result}[P\'olya-Gamma Density Ratio]
\label{propdensity}
        Fix $b = 1$. Then the pdf of P\'olya-Gamma distribution with parameters $1$ and $c$ is given as
        \begin{align*}
            f_{\mathrm{PG}}(x\mid1, c) &= 
            \cosh(c/2) \sum_{n=1}^\infty (-1)^{n-1} \frac{2n-1}{\sqrt{2\pi x^3}}\exp\left(-\frac{(2n-1)^2}{8x}-\frac{c^2}{2}x\right).
        \end{align*}
        In particular,
        \begin{align*}
            \frac{f_{\mathrm{PG}}(x\mid1, c)}{f_{\mathrm{PG}}(x\mid1, 0)} &=
            \cosh(c/2)\exp\left(-\frac{c^2}{2}x\right).
        \end{align*}
\end{result}
The derivation of Result \ref{propposterior} is given below; The full conditional posterior distribution of $\beta_h$ and $\sigma^2_h$ is 
    {\small
    \begin{align*}
        p(\beta_h, \sigma^2_h\mid \Dat^\star_h, \nu_0, s^2_0, \mathbf \Lambda_\beta) 
        &\propto
        p(\beta_h, \sigma^2_h\mid\nu_0, s^2_0, \mathbf \Lambda_\beta)
        \prod_{i=1|u_i = h}^n 
        \Nor(y_i \mid 
        \left(
        \widetilde \beta^\top_h \phi_\beta(x_i) + \widetilde \gamma^\top_h \phi_\gamma(x_i) z_i, 
        \sigma^2_h
        \right)
         \\ 
        &\propto 
        \frac{1}{\sigma_h^{\nu_0+p+2+\mid\mathcal \Dat^\star_h\mid}}
        \exp\left(-\frac{\nu_0s^2_0+\beta_h^\top \mathbf \Lambda_\beta \beta_h + \sum_{u_i = h}(Y_i - \beta_h^\top \phi(Z_i, X_i))^2}{2\sigma^2_h}\right)
    \end{align*}
    }
Due to conditional conjugacy, $\beta_h$ and $\sigma_h^2$ can be updated easily via Gibbs sampling. To derive the full conditional distribution of $b_h$, we first observe
{\small 
\begin{align*}
    p(b_h\mid\overline{\Dat}^\star_h, \mathbf \Lambda_b, \bm \eta, \bm \omega) &\propto \exp\left(-\frac{1}{2}b_h^\top \mathbf \Lambda_b b_h\right)\prod_{u_i \geq h} p(\eta_{i,h},\omega_{i,h}\mid X_i, b_h) \\
    &\propto \exp\left(-\frac{1}{2}b_h^\top \mathbf \Lambda_b b_h\right)\prod_{u_i \geq h}
    \frac{\exp(\eta_{i,h} \psi(X_i)^\top b_h)}{2\cosh(\frac{1}{2} \psi(X_i)^\top b_h)}
    f_{\mathrm{PG}}(\omega_{i,h}\mid1, \psi(X_i)^\top b_h) \\
    &\propto \exp\left(-\frac{1}{2}b_h^\top \mathbf \Lambda_b b_h\right)\prod_{u_i \geq h}
    \exp\left(\eta_{i,h} \psi(X_i)^\top b_h-\frac{1}{2}\omega_{i,h}(\psi(X_i)^\top b_h)^2\right) \\
    &
    \qquad
    \times
    f_{\mathrm{PG}}(\omega_{i,h}\mid1, 0) \\
    &\propto
    \exp\left(-\frac{1}{2}b_h^\top \mathbf \Lambda_b b_h + \sum_{u_i \geq h} \left(\eta_{i,h} \psi(X_i)^\top b_h-\frac{1}{2}\omega_{i,h}(\psi(X_i)^\top b_h)^2\right)\right). 
\end{align*}
}
The full conditional distribution of $b_h$ is normal, which greatly facilitates posterior computation.

\section{Details of Posterior Computation}
\label{appendixb}
In this section, we provide details of posterior computation.  We utilize the Markov chain Monte Carlo (MCMC) method to numerically approximate the posterior distribution. Algorithm \ref{alg:gibbs} provides details of the posterior simulation. Note that $(\beta_h, \sigma^2_h)$ can be sequentially sampled from the full conditional distribution in \eqref{eqthetaposterior} by sampling $\sigma^2_h\mid \Dat^\star_h, \nu_0, s^2_0 \sim \mathrm{IG}(\nu_h/2, \nu_hs^2_h/2)$ and then sampling $\beta_h\mid\sigma^2_h, \Dat^\star_h, \mathbf \Lambda_\beta \sim \Nor(\mathbf m_{\beta_h}, \sigma^2_h \mathbf \Lambda^{-1}_{\beta_h})$.  To further facilitate sampling of $\beta_h$ and $b_h$, we use Algorithm \ref{alg:rue} \citep{rue2001fast}.

\begin{algorithm}[!t]
    \caption{
        Posterior Simulation Implementation - Gibbs Sampler
    }
    \begin{algorithmic}[1]
    \label{alg:gibbs}
        \setstretch{1.2}
        \bigskip
        \item[] \textbf{Goal}: Sample Posterior Draws

        \STATE
        Initialize.

        \FOR{each Gibbs sampler iteration}

        \FOR{$i = 1, \ldots, n$}
            \STATE Update membership variable $u_i$ from the categorical distribution with probabilities
            \begin{align*}
                p(u_i = h \mid -) \propto w_h(X_i) \Nor\left(y_i \mid  
        \widetilde \beta^\top_h \phi_\beta(x_i) + \widetilde \gamma^\top_h \phi_\gamma(x_i) z_i, 
        \sigma^2_h
        \right).
            \end{align*}
        \ENDFOR

        \STATE
        Update $\zeta$ and $\rho_j$, $j=1, \ldots, q$ in $\mathbf \Lambda_b$ from their full conditionals. 
        \FOR{$h = 1, \ldots, H$}
            \FOR{$i$ such that $u_i \geq h$}
                \STATE Update $\eta_{i,h} = 1/2$ if $u_i = h$ and $\eta_{i,h} = -1/2$ if $u_i > h$.
                \STATE Update $\omega_{i,h} \mid - \sim \mathrm{PG}(1, \psi(X_i)^\top b_h)$.
            \ENDFOR
            \STATE Update $b_h$ from the full conditional in \eqref{eqbposterior}.
        \ENDFOR
        
        \STATE
        Update $\mathbf \Lambda_\beta$ from the full conditional in \citet{makalic2015simple}.

        \FOR{$h = 1, \ldots, H+1$}
            \STATE
            Update $\beta_h$ and $\sigma^2_h$ from the full conditional in \eqref{eqthetaposterior}.  
        \ENDFOR
        
        \ENDFOR
        
    \end{algorithmic}
\end{algorithm}

\begin{algorithm}[!t]
    \caption{
        Rue's Algorithm
    }
    \begin{algorithmic}[1]
    \label{alg:rue}
        \setstretch{1.2}
        \bigskip
        \item[] \textbf{Goal}: Sample $\beta \sim \mathcal{N}(\mathbf \Lambda^{-1}\mathbf{d}, \sigma^2\mathbf \Lambda^{-1})$, where $\mathbf{d} \in \mathbb{R}^p$ and $\mathbf \Lambda \in \mathbb{R}^{p \times p}$.

        \STATE
        Perform the Cholesky decomposition for $\mathbf \Lambda$ so that $\mathbf \Lambda = \mathbf R^\top \mathbf R$.

        \STATE
        Sample $\mathbf v \sim \mathcal{N}(0, \mathbf I_{p})$.

        \STATE
        Solve $\mathbf R \mathbf w = \mathbf v$ to obtain $\mathbf w$.

        \STATE
        Solve $\mathbf R^\top \mathbf u = \mathbf d$ to obtain $\mathbf u$.
        
        \STATE
        Solve $\mathbf R \bm\mu = \mathbf u$ to obtain $\bm\mu$.

        \STATE
        Set $\beta = \bm\mu + \sigma \mathbf w$.
        
    \end{algorithmic}
\end{algorithm}

\vskip 0.2in
\bibliography{References}

\end{document}